\documentclass[]{aastex631}

\usepackage{amsmath}
\usepackage{natbib}

\defcitealias{Baek2020}{B20}

\shorttitle{}
\shortauthors{Francis et al.}
\graphicspath{{./}{figures/}}

\begin{document}

\title{Accretion Burst Echoes as Probes of Protostellar Environments and Episodic Mass Assembly}

\correspondingauthor{Logan Francis}
\email{loganfrancis3@uvic.ca}

\author[0000-0001-8822-6327]{Logan Francis}
\affiliation{Department of Physics and Astronomy, University of Victoria, 3800 Finnerty Road, Elliot Building, Victoria, BC, V8P 5C2, Canada}
\affiliation{NRC Herzberg Astronomy and Astrophysics, 5071 West Saanich Road, Victoria, BC, V9E 2E7, Canada}

\author[0000-0002-6773-459X]{Doug Johnstone}
\affiliation{NRC Herzberg Astronomy and Astrophysics, 5071 West Saanich Road, Victoria, BC, V9E 2E7, Canada}
\affiliation{Department of Physics and Astronomy, University of Victoria, 3800 Finnerty Road, Elliot Building, Victoria, BC, V8P 5C2, Canada}

\author[0000-0003-3119-2087]{Jeong-Eun Lee}
\affiliation{School of Space Research, Kyung Hee University, 1732, Deogyeong-daero, Giheung-gu, Yongin-si, Gyeonggi-do 17104, Republic of Korea}

\author{Gregory J. Herczeg}
\affiliation{Kavli Institute for Astronomy and Astrophysics, Peking University, Yiheyuan 5, Haidian Qu, 100871 Beijing, China}
\affiliation{Department of Astronomy, Peking University, Yiheyuan 5, Haidian Qu, 100871 Beijing, China}

\author[0000-0002-7607-719X]{Feng Long}
\affiliation{Center for Astrophysics, Harvard \& Smithsonian, 60 Garden Street, Cambridge, MA 02138, USA}

\author[0000-0002-6956-0730]{Steve Mairs}
\affiliation{SOFIA Science Center, Universities Space Research Association, NASA Ames Research Center, Moffett Field, California 94035, USA}
\affiliation{East Asian Observatory, 660 N. A`oh\={o}k\={u} Place,
Hilo, Hawai`i, 96720, USA}

\author[0000-0003-1894-1880]{Carlos Contreras-Pe\~na}
\affiliation{Department of Physics and Astronomy, Seoul National University, 1 Gwanak-ro, Gwanak-gu, Seoul 08826, Republic of Korea}
\affiliation{School of Space Research, Kyung Hee University, 1732, Deogyeong-daero, Giheung-gu, Yongin-si, Gyeonggi-do 17104, Republic of Korea}

\author{Gerald Moriarty-Schieven}
\affiliation{NRC Herzberg Astronomy and Astrophysics, 5071 West Saanich Road, Victoria, BC, V9E 2E7, Canada}

\collaboration{99}{The JCMT Transient Team}



\begin{abstract}
    Protostars likely accrete material at a highly time variable rate, however, measurements of accretion variability from the youngest protostars are rare, as they are still deeply embedded within their envelopes. Sub-mm/mm observations can trace the thermal response of dust in the envelope to accretion luminosity changes, allowing variations in the accretion rate to be quantified. In this paper, we present contemporaneous sub-mm/mm light curves of variable protostars in Serpens Main, as observed by the ALMA ACA, SMA, and JCMT. The most recent outburst of EC 53 (V371 Ser), an $\sim 18$ month periodic variable, is well-sampled in the SMA and JCMT observations. The SMA light curve of EC 53 is observed to peak weeks earlier and exhibit a stronger amplitude than at the JCMT. Stochastic variations in the ACA observations are detected for SMM 10 IR with a factor $\sim 2$ greater amplitude than as seen by the JCMT. We develop a toy model of the envelope response to accretion outbursts to show EC 53's light curves are plausibly explained by the delay associated with the light travel time across the envelope and the additional dilution of the JCMT response by the incorporation of cold envelope material in the beam. The larger JCMT beam can also wash out the response to rapid variations, which may be occurring for SMM 10 IR. Our work thus provides a valuable proof of concept for the usage of sub-mm/mm observations as a probe of both the underlying accretion luminosity variations and the protostellar environment. 
\end{abstract}

\keywords{}

\section{Introduction} 
\label{sec:intro}

The infall of material from the envelopes surrounding the youngest protostars inevitably produces a circumstellar accretion disk, which serves as the primary reservoir of material for further stellar growth. Accretion through the disk is expected to be highly episodic in nature, and indeed, a wide variety of variability and accretion outburst phenomena are commonly observed in the more evolved T Tauri stars (See recent review by \citealt{Fischer2022}). Owing to their deeply embedded nature however, accretion variability in protostars is less often observed, and consequently more poorly understood. This earliest period of the stellar life cycle is where the majority of the final mass ($\gtrsim 90\%$, \citealt{Fischer2022}) is assembled, however, and up to $\sim 25\%$ of the mass may be accumulated in outbursts \citep{McKee2011,fischer2019}. Constraining accretion behaviours in protostars is thus of fundamental importance for star formation theory. 

The lack of emission at near-IR to UV wavelengths from protostars necessitates alternative diagnostics for accretion variability, as traditional measures such as emission line strength and UV-excess measurement \citep{Hartmann2016} are inaccessible. Indirectly, the clumpy structure of outflows (e.g. \citep{Plunkett2015,Jhan2022}) and the envelope chemistry \citep[e.g.][]{Jorgensen2013} can provide a fossil record of past accretion activity. Changes in accretion rate produce a directly observable response in the envelope emission at far-IR and longer wavelengths, as the shorter wavelength emission from the accretion luminosity is absorbed and re-radiated. This was first shown for simple spherically symmetric envelope models and parameterized outbursts by \cite{Johnstone2013}, while \cite{MacFarlane2019a,MacFarlane2019b} considered the observability of outbursts produced in hydrodynamical simulations of unstable disks. The response at far-IR ($\sim 100 \mu$m) wavelengths should be approximately proportional to the accretion luminosity change, whereas the sub-mm/mm response instead traces the temperature change in the envelope \citep{ContrerasPena2020}.

A variety of bright outbursts have thus been detected at mid-IR to millimeter wavelengths. The first class 0 protostar for which a strong outburst was detected was HOPS 383, which brightened by a factor of $\sim35$ at 24$\mu$m \citep{Safron2015}. Similar mid-IR outbursts from other class 0 protostars have since been found for HOPS 12, HOPS 124, \citep{Zakri2022} and V2775 Ori (HOPS 223) \citep{fischer2019}. Outbursts have also been detected towards high-mass protostars. For example, the NGC 6334-I star forming region was found from the comparison of millimeter observations to have increased in luminosity by a factor of $\sim 70$ \citep{Hunter2006,Hunter2017}, while multiple outbursts of the massive protostar M17 MIR have been observed in the mid-infrared \citep{Chen2021}, the most recent of which corresponds to a luminosity change of a factor $\sim 6$.

Although the aforementioned outbursts were all detected serendipitously, systematic efforts to quantify lower-level variability and constrain the frequency of bright outbursts from embedded protostars have recently begun. In the ongoing James Clerk Maxwell Telescope (JCMT) Transient Survey \citep{Herczeg2017}, 8 nearby ($<500$ pc) star forming regions are being monitored with a monthly or better cadence at 450 and 850 $\mu$m. An analysis of the first 4 years of Transient observations found 18 of 83 class 0 or I protostars to exhibit moderate secular variability on timescales of a few years, though the estimated mass accreted during these variations was at most a few percent of the stellar mass \citep{Lee2021}. A search for mid-IR variability in young stellar objects from the 6.5 yr Near-Earth Object Wide-field Infrared Survey Explorer (NEOWISE) All Sky Survey \citep{Mainzer2011} found $\sim 1700$ out of $\sim 5400$ protostars varying with a wide range of behaviours, with the youngest protostars the most likely to exhibit variability \cite{Park2021}. While mid-IR variability can also be attributed to changes in extinction and viewing geometry, a comparison of varying sources detected in both the Transient Survey and NEO(WISE) observations found that the sub-mm variable sources typically showed similar variability in the mid-IR. Furthermore, these sources constituted about 22\% of the overall sample, suggesting accretion variability is responsible for the mid-IR flux changes in many cases \cite{ContrerasPena2020}.

While these large monitoring campaigns are important for establishing the distribution of accretion variability events, their low angular resolution precludes a detailed investigation of how the envelope structure may influence the response to accretion luminosity variations. Modeling of the envelope response predicts that higher resolution observations may be more sensitive to accretion rate changes, as the change in brightness of the outer envelope is likely to be dominated by heating from the interstellar radiation field \citep{Johnstone2013}. The details of the envelope structure, including the likely presence of outflow cavities and dust sublimation fronts, may also impact the observed response \citep[][hereafter B20]{Baek2020}. High resolution monitoring of embedded protostars with millimeter wavelength facilities should thus be able to probe the envelope structure, in addition to identifying variability behaviour. In this paper, we thus analyze observations from the Atacama Large Millimeter/Submillimeter Array (ALMA) and Submillimeter Array (SMA) of variable protostars in Serpens Main monitored by the JCMT Transient Survey. Particular emphasis is placed on EC 53 (V371 Ser), which exhibits $\sim 18$ month periodic accretion bursts, the most recent of which was observed at high cadence with the SMA and JCMT. We interpret the light curves from our monitoring programs using a simple toy model of the propagation of accretion bursts through the envelope, which we also use to further explore how the properties of the envelope and observational setup affect the burst response.

 The remainder of this paper is organized as follows: In Section \ref{sec:obs}, we describe our variable protostar targets and the ALMA, SMA, and JCMT observations, while in Section \ref{sec:dr}, we provide the details of our data reduction and present light curves of our targets. In Section \ref{sec:ec53_toy_modeling}, we develop a toy model to interpret the SMA and JCMT observations of EC 53, followed in Section \ref{sec:further_modeling} by further exploration with the toy model on the observation of generic accretion bursts. We then a discuss our results and observational/modeling caveats in Section \ref{sec:disc}, and finish with a brief summary of our major conclusions in Section \ref{sec:conc}.

\section{Observations and Standard Calibration}
\label{sec:obs}

\begin{figure}
    \centering
    \includegraphics{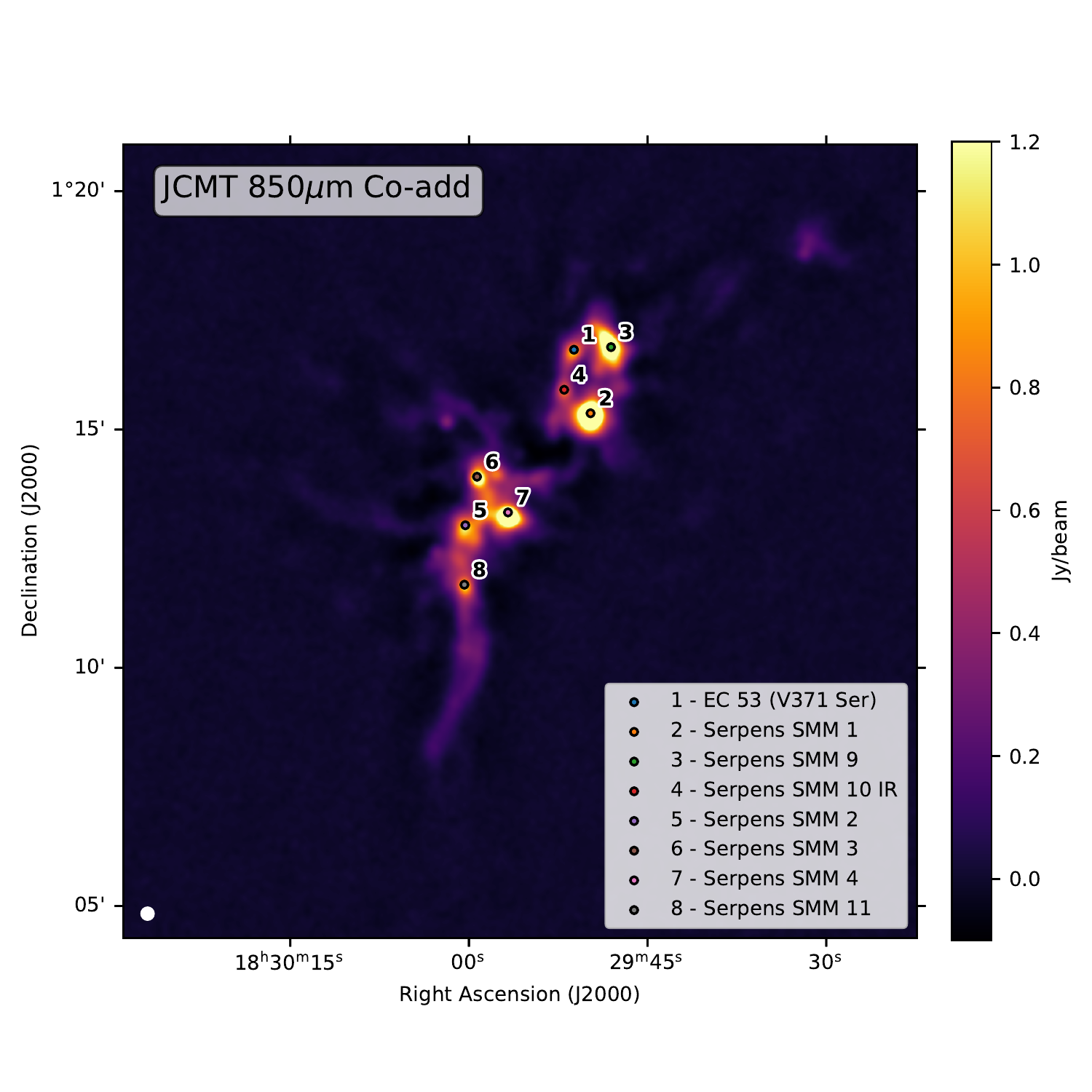}
    \caption{A subsection of the Serpens Main star-forming region monitored by the JCMT at 850 \micron, annotated to show the positions of the 4 variable protostars and 4 stable calibrator sources. The map combines all 81 epochs observed through March 14, 2022. The full width at half maximum of the beam is shown by the white circle.}
    \label{fig:finder_chart}
\end{figure}

In this section, we describe the details of the ALMA Atacama Compact Array (ACA), Submillimeter Array (SMA), and James Clerk Maxwell Telescope (JCMT) observations used in this paper. The targets of our observations are all located in the Serpens Main star forming region, at a distance of 436 pc \citep{Ortiz-Leon2017,herczeg2019}. An overview of Serpens Main as it appears in a co-add of the Tranient Survey observations is shown in Figure \ref{fig:finder_chart}. 

The ACA observations monitor the thermal dust continuum of 3 known variable protostars (EC 53/V371 Ser, Serpens SMM1, Serpens SMM10 IR) and 5 YSOs (young stellar objects) intended for use as stable calibrators which were identified by the JCMT Transient Survey \citep{Lee2021}. The coordinates and basic details of each target are listed in Table \ref{tab:targets}. The strongest variations in the Transient Survey are seen from EC53 (V371 Ser), a class I protostar which undergoes periodic outbursts lasting $\sim 6$ weeks, followed by a slow decline in brightness until the next outburst every $\sim$ 18 months \citep{Yoo2017}. Modeling of multi-wavelength observations of EC 53 associates a reddening of the protostar just prior to the burst with the build-up and subsequent draining of material in the inner disk \citep{Lee2020yh}. Radiative transfer simulations of EC 53's envelope and comparison with the 850 $\mu$m Transient Survey observations suggest that the accretion luminosity increases by a factor of $\sim 3.3$ during each outburst \citepalias{Baek2020}, although the strength and duration between bursts varies to some degree \citep{Lee2020yh}. Our other targets selected from the Transient Survey are Serpens SMM 1, a intermediate mass protostar which shows a steady rise in brightness over several year timescales, and Serpens SMM 10 IR, which exhibits stochastic variations in brightness \citep{Johnstone2018,Lee2021}. The other five YSOs were intended to be monitored to provide a stable reference flux for relative calibration (Section \ref{sec:rel_calibration}), however, two were found to be unsuitable for this purpose. SMM 9 is now found to be moderately variable in the Transient Survey \citep{Lee2021}, while SMM 2 is too faint and extended at the resolution of the ACA to provide a relative calibration, as discussed by \cite{Francis2020}. The remaining three YSOs (SMM 3, SMM 4, and SMM 11) are thus the only ones used for relative calibration, and their brightness has remained stable (RMS $<2\%$) over the lifetime of the JCMT Transient Survey. 

The SMA observations were taken with the specific goal of capturing the 2021 outburst of EC 53 with a high cadence, and thus only targeted EC 53 and the 3 stable calibrator sources used by the ACA (see Table \ref{tab:targets}).

\begin{deluxetable*}{lhllcc}[h]
\label{tab:targets}
\tablecaption{Science Targets and YSO Calibrators}
\tablehead{\colhead{Name} & \nocolhead{ALMA Name} & \colhead{Other mm Source Names} & \colhead{ACA field center (ICRS)}\tablenotemark{a} & \colhead{SMA Target?} & \colhead{Target Type}}
\startdata
EC 53 (V371 Ser)         &  Serpens\_Main\_850\_02 & Ser-emb 21          & 18:29:51.18 +01:16:40.4  & Y & V \\
Serpens SMM 1            &  Serpens\_Main\_850\_00 & Ser-emb 6, FIRS1    & 18:29:49.79 +01:15:20.4  & N & V \\
Serpens SMM 9            &  Serpens\_Main\_850\_01 & Ser-emb 8, SH2-68N  & 18:29:48.07 +01:16:43.7  & N & V\tablenotemark{b} \\
Serpens SMM 10 IR        &  Serpens\_Main\_850\_03 & Ser-emb 12          & 18:29:52.00 +01:15:50.0  & N & V \\
Serpens SMM 2            &  Serpens\_Main\_850\_10 & Ser-emb 4(N)       & 18:30:00.30 +01:12:59.4  & N & S\tablenotemark{c} \\ 
Serpens SMM 3            &  Serpens\_Main\_850\_09 &         -           & 18:29:59.32 +01:14:00.5  & Y & S \\
Serpens SMM 4            &  Serpens\_Main\_850\_08 &         -           & 18:29:56.72 +01:13:15.6  & Y & S \\ 
Serpens SMM 11           &  Serpens\_Main\_850\_11 &         -           & 18:30:00.38 +01:11:44.6  & Y & S 
\enddata
\tablecomments{V=JCMT Variable; S=JCMT Stable, intended to be used for calibration.
\tablenotetext{a}{Our shared SMA and ACA targets have the same field center.}
\tablenotetext{b}{Serpens SMM 9 was originally identified as stable by the JCMT Transient Survey \citep{Johnstone2018} but has since been found to exhibit moderate variability \citep{Lee2021}.
\tablenotetext{c}{Serpens SMM 2 is bright and stable in the JCMT observations, but is too faint and extended at the resolution of the ACA to be used as a calibrator.}}
}
\end{deluxetable*}

\subsection{ALMA ACA 850 $\mu$m Observations}
\label{ssec:alma_obs}

Our ALMA programs (2018.1.00917.S, 2019.1.00475.S, PI: Logan Francis) observe the targets in Table \ref{tab:targets} using the ACA, also known as the Morita Array, a sub-array of ALMA consisting of twelve 7m diameter antennas in a fixed configuration. We acquired 10 epochs of observations over 3 years, typically taken 1-3 months apart, except for a $\sim$1.5 year gap between the first 7 and last 3 due to the shutdown of ALMA in 2020. The details of the observations for each epoch are summarized in Table \ref{tab:aca_obs}. For the first 7 epochs, the ACA correlator was configured in time division mode using the default Band 7 continuum settings, which provides an integration time of 1.01s and sets the local oscillator frequency to 343.5 GHz with four spectral windows centered at 336.5 GHz, 338.5 GHz, 348.5 GHz, and 350.5 GHz. Each spectral window provides 1.875 GHz of bandwidth across 128 channels, for a total bandwidth of 7.5 GHz. For the last 3 epochs, we requested the frequency division mode of the correlator with improved spectral resolution for better emission line characterization. Specifically, in these epochs the spectral windows have the same bandwidth and frequency center, but 2048 channels across 1.875 GHz of bandwidth, the trade-off being a longer integration time of 10.1s. The typical resolution of our observations is $\sim 4\arcsec$. Our continuum sensitivity is typically $\lesssim 1$ mJy per epoch, in Table \ref{tab:aca_obs} we also provide values of the sensitivity for individual epochs estimated using the visibility weights with the \texttt{CASA} task \texttt{apparentsens}.

Our visibility data for all epochs was calibrated in \texttt{CASA}\footnote{\url{https://casa.nrao.edu/}} 5.6.1 \citep{McMullin2007} using the ALMA pipeline. For each target, we extracted the continuum by creating dirty image cubes, then sigma-clipping the spectra measured in a 3\arcsec\ diameter aperture centered on the brightest source to determine the line free channels.  To produce deep and high fidelity images of our targets (Figure \ref{fig:aca_deep_gallery}), we first combined the data from every epoch and brought the amplitude scales of the visibilities into agreement using the relative calibration factors discussed in Section \ref{sec:rel_calibration}, then created images with the \texttt{tclean} task in \texttt{CASA} using the multifrequency synthesis deconvolution and a Briggs robust weighting of 0.5. We performed phase-only self-calibration for each epoch and all of our targets except Serpens SMM 2, which was too faint to obtain useful gain solutions. Self-calibration can slightly increase the measured flux of our targets by reducing the effect of phase decorrelation and is important for accurate determination of the average flux scale, but does not significantly affect the relative calibration accuracy \citep{Francis2020}. Three rounds of self-calibration were applied using solution intervals of a scan length, 20.2s, and 5.05s (first 7 epochs) or 10.1s (last 3 epochs).

\begin{deluxetable}{cccccccc}
\label{tab:aca_obs}
\tablecaption{ACA Observations}
\tablehead{\colhead{Date\tablenotemark{a}} & \colhead{Antennas} &\colhead{$uv$ range (m)} & \colhead{Beam Size (\arcsec)\tablenotemark{b}} & \colhead{Sensitivity (mJy)} &\colhead{Spectral Setup\tablenotemark{c}} & \colhead{Calibrators\tablenotemark{d}}}
\startdata
2018-08-14  & 10 & 7.0 - 45.1 & 5.8 $\times$ 2.7 & 0.63 & A & J1924-2914, J1851+0035 \\ 
2019-03-06  & 11 & 8.1 - 44.2 & 5.3 $\times$ 2.3 & 0.67 & A & J1751+0939, J1743-0350 \\ 
2019-04-07  & 11 & 7.8 - 47.2 & 5.8 $\times$ 2.2 & 0.52 & A & J1517-2422, J1751+0939 \\ 
2019-05-18  & 11 & 8.7 - 44.9 & 4.9 $\times$ 2.8 & 0.50 & A & J1924-2914, J1743-0350 \\ 
2019-08-04  & 9  & 7.3 - 44.9 & 4.8 $\times$ 2.9 & 0.41 & A & J1924-2914, J1743-0350 \\ 
2019-09-20  & 9  & 7.3 - 45.4 & 6.4 $\times$ 2.8 & 0.51 & A & J1924-2914, J1851+0035 \\ 
2019-10-29  & 11 & 7.0 - 44.7 & 5.2 $\times$ 2.9 & 0.43 & A & J1924-2914, J1851+0035 \\
2021-07-01  & 8  & 8.5 - 42.6 & 5.4 $\times$ 2.6 & 0.63 & B & J1924-2914, J1851+0035 \\
2021-08-02  & 8  & 8.2 - 42.0 & 5.5 $\times$ 3.5 & 0.95 & B & J1924-2914, J1851+0035 \\
2021-09-05  & 8  & 7.2 - 44.6 & 5.8 $\times$ 3.4 & 0.62 & B & J1924-2914, J1851+0035 \\
\enddata
\tablecomments{
\tablenotetext{a}{The dates provided are at the start date of each track in UTC time in year-month day format.}
\tablenotetext{b}{Beam size is the full width at half maximum of the major $\times$ minor axes of the restoring beam in tclean.}
\tablenotetext{c}{Our two spectral setups both have a local oscillator frequency of 343.5 GHz and four 2GHz wide spectral windows centered at 336.5 GHz, 338.5 GHz, 348.5 GHz, and 350.5 GHz. The correlator mode differs, however, and the spectral resolution and integration time for each setup are:\\
A: Time division mode, integration time 1.01s, 128 channels. \\
B: Frequency division mode, integration time 10.1s, 2048 channels \\}
\tablenotetext{d}{The sources listed are in the order: flux and bandpass calibrator, gain calibrator. }
}
\end{deluxetable}

\subsection{SMA 1.3 mm Observations}
\label{ssec:sma_obs}

Our observations with the SMA, an 8 element interferometer with 6m dishes, consist of 16 tracks (2020B-S044, 2021A-S056, PI: Logan Francis) used to simultaneously monitor EC 53 during its 2021 accretion outburst and 3 additional YSO calibrators (Table \ref{tab:targets}). The details of each track are summarized in Table \ref{tab:sma_obs}. Our observations were nominally requested to use either the compact or subcompact array configuration, and the 230/240 GHz RxA/RxB receiver combination with local oscillator tunings of 232.5 GHz and 244.5 GHz; however, to facilitate scheduling some flexibility was allowed, and any local oscillator tuning of the receivers within $238.5\pm10$ GHz was permitted for all but the first and final epochs. The typical beam size is $\sim 3\arcsec$, while the continuum sensitivity is $\lesssim 1$ mJy. We estimate the continuum sensitivity for each epoch in the same manner as for the ACA observations, the values of which are provided in Table \ref{tab:sma_obs}. The observations were performed using the upgraded SWARM correlator \citep{Primiani2016}, which simultaneously processes data from both receivers. The centers of the upper and lower sideband of each receiver are separated from the local oscillator frequency by $\pm10$ GHz; each sideband is divided into 6 spectral windows with 2GHz bandwidth and 140kHz channels each, providing a total processed bandwidth of up to 48 GHz from both receivers. Several initial epochs were taken in early 2021 to establish the pre-outburst flux of EC 53; after the outburst was detected to have begun at the SMA and JCMT in mid-April, an approximate 10 day cadence for the remaining observations during the EC 53 rise was requested. The compact configuration of the SMA was used for all but the final observation, which used the sub-compact configuration instead. For most epochs, only 6 of the 8 SMA antennas were available. 

Data calibration was performed using standard SMA procedures with the \texttt{MIR} software\footnote{\url{https://lweb.cfa.harvard.edu/~cqi/mircook.html}}, which we briefly describe here. The raw data was first binned down by a factor of 8 in spectral resolution using the SMARechunker tool\footnote{\url{https://github.com/Smithsonian/SMARechunker}}. This binning reduces the computational resources needed for reduction while still allowing good differentiation between emission lines and continuum. Baseline correction was performed if needed, and periods of  noisy/corrupted data were flagged out. Spectral spikes in the data were corrected, followed by system temperature correction and bandpass calibration. Flux calibration was performed by measuring the gain calibrator flux using the brightest solar system object available during the track, then transferring the flux of the gain calibrator to the science targets during phase and amplitude calibration. The final data for each science target was exported to \texttt{CASA} measurement set format. 

Deep images of each SMA target were constructed by the same procedure as for the ACA observations (Section \ref{ssec:alma_obs}), by first extracting the continuum from the dirty cubes, then applying relative calibration factors for each SMA epoch (Section \ref{sec:rel_calibration}) and imaging the combined data from all 16 epochs using \texttt{tclean} with a Briggs robust value of 0.5. Due to the lower S/N on individual integrations with the SMA compared with the ACA, we did not apply self-calibration to our SMA observations. The final deep images are shown in Figure \ref{fig:sma_deep_gallery} in the Appendix. The visual appearance of our shared targets at the SMA and and ACA are similar, however, the SMA resolution and imaging fidelity is somewhat better than the ACA owing to the improved uv-coverage provided by the more optimal location of the SMA for targets near the celestial equator, additional epochs, and longer observation lengths, typically 5-8\,hrs for each SMA track versus $\sim$1 hr for the ACA. Additionally, the primary beam of the SMA is much larger than the ACA ($\sim 55\arcsec$ vs $\sim 30\arcsec$), providing better sensitivity to sources away from the field center.

\begin{deluxetable}{ccccccc}
\tablecaption{SMA Observations}
\label{tab:sma_obs}
\tablehead{\colhead{Date} & \colhead{Antennas} &\colhead{$uv$ range (m)} & \colhead{Beam Size (\arcsec)} & \colhead{Sensitivity (mJy)} & \colhead{Spectral Setup\tablenotemark{a}} & 
\colhead{Calibrators\tablenotemark{b}}}
\startdata
2021-02-19 & 6 & 16.2 - 75.3 & 2.8 $\times$ 2.3 & 0.94 & A & Vesta,      3c279  \\
2021-03-05 & 7 & 11.6 - 74.9 & 3.1 $\times$ 2.7 & 0.38 & B & Vesta,      3c279  \\
2021-03-19 & 8 & 14.6 - 74.8 & 2.9 $\times$ 2.7 & 0.34 & B & Vesta,      3c279  \\
2021-04-02 & 7 & 14.5 - 75.8 & 3.0 $\times$ 2.6 & 0.39 & B & MWC349a,    3c279  \\
2021-04-18 & 6 & 16.4 - 72.0 & 2.9 $\times$ 2.5 & 0.44 & C & MWC349a,    3c279  \\
2021-05-29 & 6 & 13.9 - 75.9 & 3.2 $\times$ 2.4 & 0.32 & B & Callisto,   3c279  \\
2021-06-09 & 6 & 15.9 - 76.0 & 3.0 $\times$ 2.5 & 0.27 & B & Ganymede,   3c279  \\
2021-06-18 & 6 & 15.2 - 76.2 & 3.0 $\times$ 2.4 & 0.47 & B & MWC349a,    3c279  \\
2021-06-27 & 6 & 16.6 - 75.4 & 2.9 $\times$ 2.5 & 0.48 & B & MWC349a,    3c84   \\
2021-07-06 & 6 & 14.5 - 76.2 & 2.9 $\times$ 2.4 & 0.65 & B & Titan,      3c84   \\
2021-07-18 & 6 & 14.5 - 75.8 & 2.8 $\times$ 2.4 & 0.31 & B & Callisto,   3c84   \\
2021-08-03 & 5 & 10.4 - 59.5 & 5.1 $\times$ 3.0 & 0.55 & B & Callisto,   3c84   \\
2021-08-11 & 6 & 14.7 - 74.6 & 2.8 $\times$ 2.3 & 0.61 & A & Vesta,      3c279  \\
2021-08-12 & 6 & 14.5 - 76.0 & 2.8 $\times$ 2.5 & 0.84 & A & Callisto,   3c279  \\
2021-08-18 & 6 & 14.5 - 69.1 & 5.2 $\times$ 2.4 & 0.36 & A & Callisto,   BL Lac \\
2021-09-21 & 7 & 8.1 - 69.0  & 3.5 $\times$ 2.7 & 0.31 & A & Callisto,   BL Lac \\
\enddata
\tablecomments{
The date and beam size are formatted in the same manner as Table \ref{tab:aca_obs}.
\tablenotetext{a}{Our three spectral setups are as follows:\\
A: Our requested default tuning, used for first and last 4 epochs. LO Tunings: Rx230 = 232.500 GHz, Rx240 = 244.500 GHz\\
B: A ``Standard" SMA continuum tuning, used for most epochs. LO Tunings: Rx230 = Rx240 = 225.538 GHz\\
C: Unique tuning used for 5th epoch only. LO Tunings: Rx230 = Rx240 = 215.100 GHz}
\tablenotetext{b}{The sources listed are in the order: flux calibrator, bandpass calibrator. We also use 1743-038 for gain calibration in every epoch.}
}
\end{deluxetable}

\subsection{JCMT 850 $\mu$m Observations}
\label{ssec:jcmt_obs}

The JCMT Transient Survey \citep[M16AL001, M20AL007;][]{Herczeg2017,Lee2021} monitors eight Gould Belt star forming regions in the 450 and 850 $\mu$m continuum bands using the SCUBA-2 instrument \citep{Holland2013}. Each map is observed to a uniform depth $\sim$ 12 mJy per beam at 850 $\mu$m for ease of comparison across epochs \citep[see][]{Mairs2017Cal}. The weather sensitivity of the 450 $\mu$m observations, however, yields a large, order of magnitude, range in observation depth, making the shorter wavelength monitoring significantly more complicated (Mairs et al.\ in preparation). In this paper we concentrate on the 850 $\mu$m measurements and only briefly discuss the 450 $\mu$m light curve for EC 53 (Section \ref{ssec:caveats}). 

Each star forming region is observed with a 30 arcminute diameter circular footprint, using the Pong 1800 scanning mode \citep{kackley2010}. For Serpens Main the map is centered at (R.A., Decl.) = (18:29:49,+01:15:20, J2000). All of our ACA and SMA targets are in this region, which when visible has been observed at a monthly or better cadence since February 2016, and with an approximate one week cadence during the 2021 outburst of EC 53. The effective beam size at the JCMT is 14.4\arcsec\ at 850 $\mu$m and 10.0\arcsec\ at 450 $\mu$m \citep{mairs2021}; however, to better measure the peak fluxes and account for variations in the telescope beam between epochs, the maps are convolved to 15.6\arcsec\ and 10.8\arcsec\ at 850 and 450 $\mu$m, respectively.

The enhanced, second generation, JCMT relative flux calibration strategy is described in detail by {Mairs et al.\ (in preparation)} and offers a small improvement over the original relative calibration scheme \citep{Mairs2017Cal}. This calibration method is similar to that described below for the ACA and SMA. However, given the large number of JCMT epochs to date (81 as of March 14, 2022) every non-robust variable in the region is included as an epoch calibrator, weighted by its flux uncertainty across all epochs. An iterative approach is used to determine the best relative flux calibration for each epoch. For Serpens Main, the epoch calibration uncertainty at 850 $\mu$m is measured to be $1.5\%$. The expected flux uncertainty for a source in a given epoch includes both the map noise, 12 mJy\,bm$^{-1}$, and the flux calibration uncertainty applied to the source, added in quadrature. 
 
A deep image of Serpens Main is produced by stacking all the JCMT epochs. At 850 $\mu$m this results in an image with a noise $\sim$ 1.1 mJy\,bm$^{-1}$. The central region of Serpens Main in presented in Figure \ref{fig:finder_chart} along with the locations of the ACA and SMA targets. The four stable calibrator sources (Table \ref{tab:targets}) have measured fractional RMS brightness values across the 81 monitored epochs of 1.5\%, with the exception of SMM 2 (1.8\%).

\section{ACA and SMA Data Reduction}
\label{sec:dr}
\subsection{Relative Flux Calibration}
\label{sec:rel_calibration}

Flux calibration at submm/mm wavelengths is typically  accurate to 10-20\%, as calibration is usually performed with solar system objects or quasars, which have uncertain fluxes owing to difficulties in modeling the emission and in the inherent time variability \citep[see for example,][]{Francis2020}.  To achieve high precision measurements of protostellar variability, our ACA and SMA observations use a relative flux calibration strategy pioneered for the JCMT Transient Survey \citep{Mairs2017Cal}, whereby a number of bright and stable calibrator YSOs are used to correct the flux scales at each epoch. This calibration strategy, applied to interferometric observations, is described extensively by \citet{Francis2020}, where it was used to  independently assess the calibration accuracy of the ALMA pipeline. Here, we provide an overview of this method as applied to our ACA and SMA observations.

Our ACA and SMA relative calibrations both use SMM 3, SMM 4, and SMM 11 (Table \ref{tab:targets}). Although the ACA observed two additional sources that were initially classified as stable in the JCMT Transient Survey, SMM 2 and SMM 9, we do not include them in the calibration. While SMM2 is bright and stable in the Transient Survey owing to the large $\sim15\arcsec$ JCMT beam, at the $\sim4\arcsec$ ACA resolution it is too faint and extended to obtain a useful calibration. Serpens SMM 9 was originally identified as stable in the JCMT Transient Survey \citep{Johnstone2018} but after four years of monitoring was found to exhibit moderate variability \citep{Lee2021}, and we thus also exclude it as a calibrator. 

To obtain a relative calibration, we first perform a $\chi^2$ fit of a point source model in the $uv$-plane to each calibrator and epoch using the \texttt{lmfit} library\footnote{\url{https://lmfit.github.io/lmfit-py/}}. For the SMA data, we additionally fit visibility data from receiver (Rx230 and Rx240) independently to take into account possible systematic differences between instruments. Our choice of a point source model is reasonable, as at the resolution of the ACA and SMA, our chosen calibrator sources are approximately point-like with some much fainter extended emission (see Figures \ref{fig:aca_deep_gallery} and \ref{fig:sma_deep_gallery}), and moreover, we find it sufficient for precise relative calibration. Performing our model fitting in the $uv$-plane also renders error analysis more tractable by avoiding the uncertainties common to image reconstruction algorithms.

\begin{figure}[htb]
    \centering
    \includegraphics[width=0.85\textwidth]{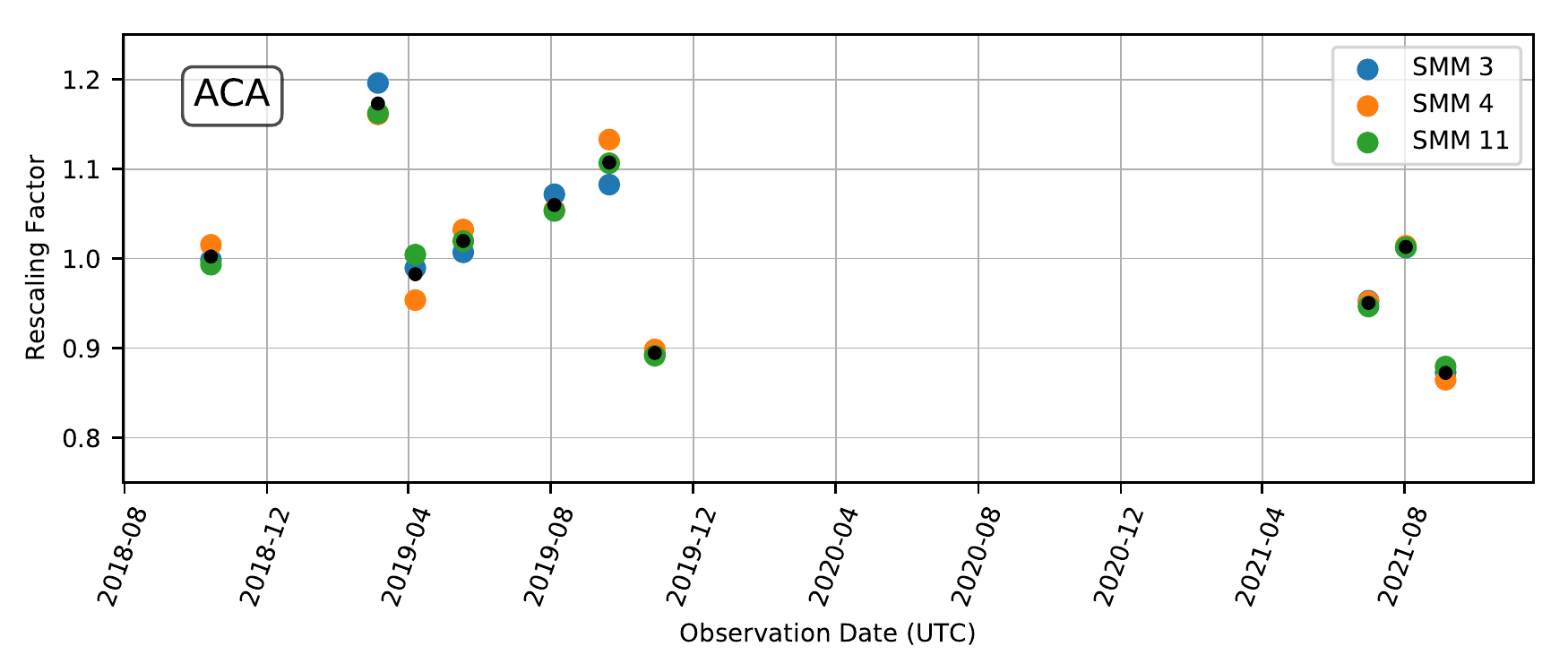}
    \caption{ACA Relative Calibration. Rescaling factors versus date for each of the three YSO calibrators (colored circles) and the mean relative flux scaling factor (black circle).}
    \label{fig:aca_calibration}
\end{figure}

\begin{figure}[htb]
    \centering
    \includegraphics[width=0.85\textwidth]{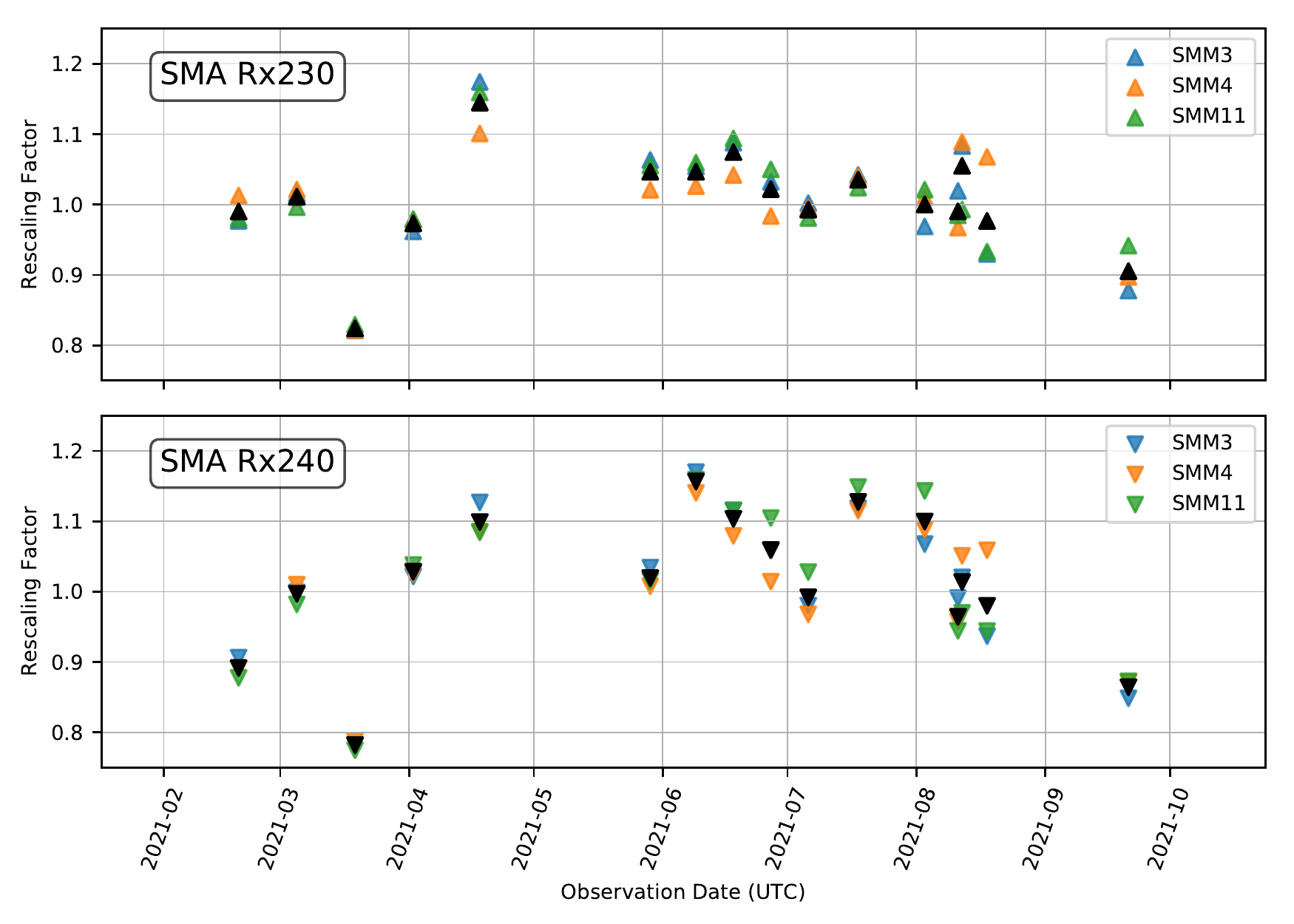}
    \caption{SMA Relative Calibration. Rescaling factors versus date for each of the three YSO calibrators (colored triangles) and the mean relative flux scaling factor (black triangle). {\it Top panel}: Rx230. {\it Bottom panel}: Rx240.}
    \label{fig:sma_calibration}
\end{figure}

For our ACA observations, we proceed by using our point source fits to calculate the mean flux across all epochs for each calibrator. For each epoch and calibrator, we determine the ratio required to correct its flux to the mean across epochs. Then, we calculate a relative flux calibration factor for each epoch by taking the average across calibrators. Our procedure for the SMA data is the same, but performed independently for each receiver. The resulting relative flux calibration factors for all of our ACA and SMA epochs are shown by the black symbols in Figures \ref{fig:aca_calibration} and \ref{fig:sma_calibration} respectively. By examining the distribution of all mean correction factors (colored symbols) across calibrators, we estimate the uncertainty in our relative flux calibration factors to be $3\%$ for the ACA, the same accuracy achieved using the first 7 epochs by \cite{Francis2020}, and 2.5\% (per receiver) for the SMA. For both telescopes, the range of the mean correction factors is $\sim20$\%, while the standard deviation is 7-10\%, as expected. Our relative calibration method thus allows us to construct light curves for our targets that are sensitive to much smaller relative variations in flux than would otherwise be possible with conventional mm flux calibration strategies.

\subsection{Calibrated Flux Measurements of Known JCMT Variables}

In order to accurately measure the flux of our variable targets (EC 53, SMM 1, SMM 10) in the $uv$-plane at the scale of the beam size in our ACA and SMA observations, we model and remove the extended emission (i.e., the outer envelope and outflow cavity edges), which we assume to exhibit little to no variability. This approach avoids systematic errors resulting from biasing the flux of the central source towards higher values when including the high amplitude visibilities on the shortest baselines. For the purposes of our relative calibration, we have found that point source models alone are sufficient for delivering a high calibration accuracy. 

We construct a model for the extended emission by separating out the bright central source in the deconvolved deep images of our targets which combine data from all epochs (see Figures \ref{fig:aca_deep_gallery} and \ref{fig:sma_deep_gallery}). This is carried out by examining the distribution of flux over all pixels in the \texttt{.model} files produced by \texttt{tclean}, that contain the collection of point sources representing real emission before convolution by the clean beam. We first remove any model pixels with negative flux which may have been introduced during the cleaning process, and then identify the bright outlier pixels representing the central source by assigning a flux cutoff threshold using a multiple of the standard deviation of the pixel distribution, typically between 15-60$\sigma$. Because the default \texttt{tclean} algorithm represents all emission using point sources, models of extended structures usually appear patchy or speckled. We thus smooth the extended structure with a small circular Gaussian with approximately half the beam size: $1.5\arcsec$ for the SMA and $2\arcsec$ for the ACA. Finally, we transform our smoothed images of the extended emission to the $uv$-plane using the \texttt{galario} library. 

We fit the visibility data without relative calibration for each target and epoch (and for the SMA, each receiver) using a model consisting of the extended emission model plus a point source representing the bright central source. The extended emission model is matched to the flux scale of each epoch using the relative flux calibration factors and held fixed, while the amplitude of the point source is allowed to vary. Although the extended emission may also exhibit variability as changes in accretion luminosity propagate to the outer envelope, we assume it's flux to be constant as it is faint in comparison to the central source. We perform our $\chi^2$ fit of the models in the $uv$-plane using \texttt{lmfit}, and then use the relative flux calibration factors to correct the resulting point source flux scales.

\subsection{Light Curves of Known JCMT Variables}
\label{sec:light_curves}

\begin{figure}[t]
    \centering
    \includegraphics[width=0.8\textwidth]{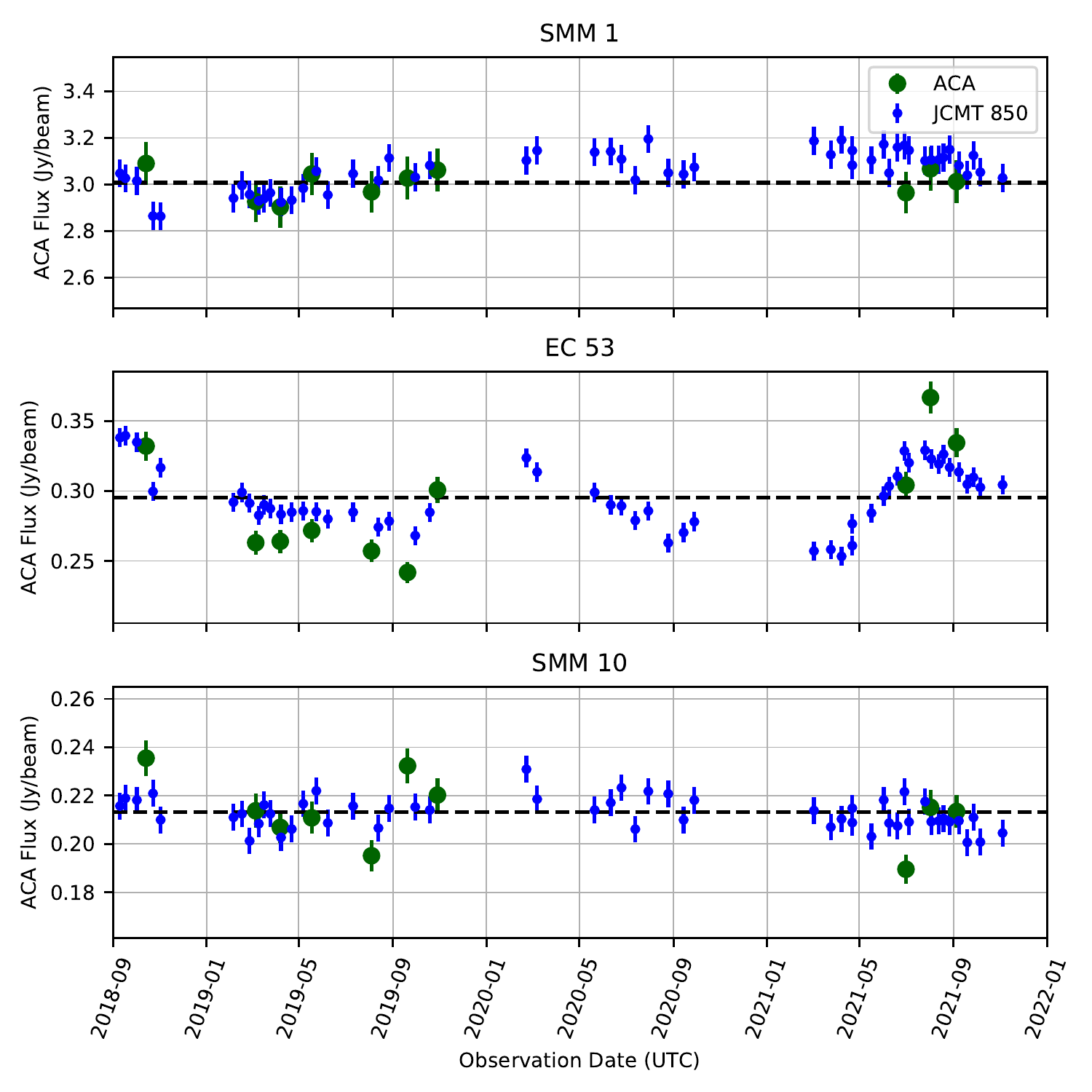}
    \caption{Light curves for the variable targets observed with the ACA (large green circles) and contemporaneous JCMT 850$\mu$m observations (small blue circles). The ACA Flux measurements are shown after relative calibration (Section \ref{sec:rel_calibration}) has been applied. The JCMT data for each source has been scaled such that the mean light curve flux is equal to the mean of the ACA data (dashed black line).}
    \label{fig:aca_lc}
\end{figure}

\begin{figure}[htb]
    \centering
    \includegraphics[]{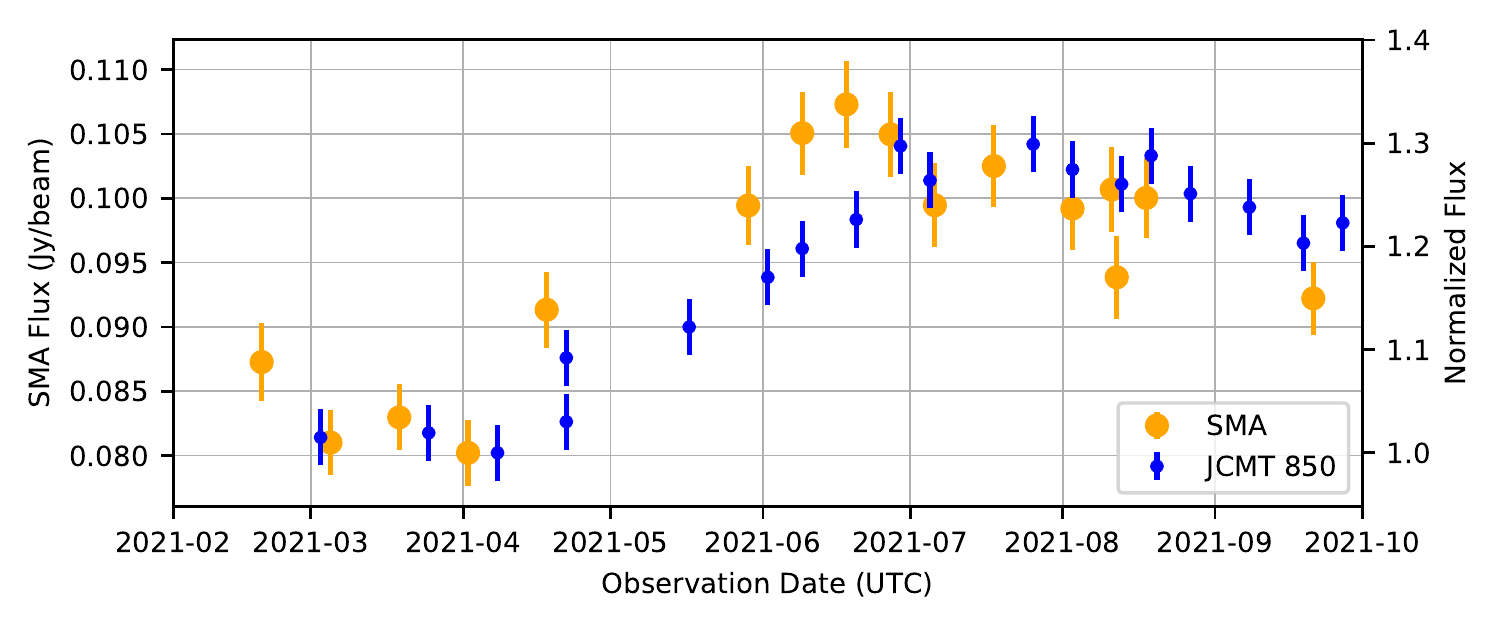}
    \caption{Light curves for EC 53 as observed during the 2021 outburst by the SMA (large orange circles) and JCMT (small blue circles). The left-side scale shows the SMA flux with our relative flux calibration applied (see Section \ref{sec:rel_calibration}). The right-side scale shows the SMA and JCMT Flux, normalized to the minimum observed flux of EC 53 on 2021-04-02 for the SMA and 2021-04-08 for the JCMT.}
    \label{fig:sma_lc}
\end{figure}

The final light curves for all the variable targets observed by the ACA are shown in Figure \ref{fig:aca_lc}. The error bars include both the source measurement uncertainty as well as the 3\,\% relative calibration uncertainty. For comparison, we overlay the contemporaneous JCMT 850 $\mu$m light curves of the same targets, with the data scaled such that the mean of the JCMT light curve is equal to the mean at the ACA, indicated by the dashed horizontal line. Similarly, we show the light curves from the SMA and JCMT for EC 53 during the 2021 outburst in Figure \ref{fig:sma_lc}. The average SMA flux across receivers is shown on the left dependent axis. On the right dependent axis, the normalized flux for the SMA and JCMT is shown, which is calculated by normalizing the data to the minimum observed flux during the outburst. These minima occur on 2021-04-02 and 2021-04-08 for the SMA and JCMT respectively.

At the JCMT, Serpens SMM 1 has steadily brightened from early 2019 to Spring 2021, followed by a somewhat steeper decay till the end of 2021. Our ACA light curve of SMM 1 is consistent with a constant flux within observational uncertainties. In the most recent ACA observations, we either caught the decline earlier than at the JCMT owing to the light travel time through the envelope (see Section \ref{sec:further_modeling}), or have simply not significantly detected the rise in flux given the relatively larger uncertainty in the ACA flux (3\,\%) compared to the JCMT ($<2$\,\%). 

In the ACA observations of EC 53, we detect a decline in amplitude from late 2018 until the beginning of its outburst in late 2019, and a rise and fall in flux over 3 months of observations roughly centered on the burst peak as seen by the JCMT. The overall amplitude of the EC 53 light curve appears twice as strong as that seen by the JCMT when comparing the 2019 minimum flux and peak in Summer 2021. The individual outbursts of EC 53 vary in amplitude \citep{Lee2020yh}, so the lack of data near the 2019 outburst peak and 2021 outburst minimum precludes a robust comparison of the observed response to individual bursts between the ACA and JCMT (see Section \ref{ssec:caveats}). 

At the SMA, the outburst of EC 53 is observed to peak weeks earlier and decline more rapidly than in the JCMT observations. Furthermore, the amplitude of the outburst in the SMA observations $\sim$35\%, is stronger than for the JCMT, $\sim$30\%, though this difference is only somewhat larger than the uncertainty in the flux measurements of a few percent.

The variability of SMM 10 at the JCMT is characterized by a many-month brightening and fading event \citep{Lee2021} that peaked in late 2017 and early 2018, before the ACA observations. Since that time SMM 10 has remained a stochastic source, where the standard deviation of the flux, $3\%$, exceeds the variations expected from noise and calibration errors, $\sim$1.5\%. Curiously, in the contemporaneous ACA observations we detect much stronger evidence of variability: the standard deviation of the ACA data is 0.014 Jy/beam or $\sim10$\,\%, a factor of 2 greater than that seen for the JCMT data. 

This variability is uncorrelated with the JCMT light curve but exhibits a similar stochastic behaviour.

\section{Modeling the SMA AND JCMT Light Curves of EC 53}
\label{sec:ec53_toy_modeling}

The amplitude and structure of the ACA and SMA light curves in comparison with the simultaneous JCMT observations (Figures \ref{fig:aca_lc} and \ref{fig:sma_lc}) suggests that the spatial resolution of the observations, and thus the physical envelope scales probed, play an important role in determining the observed light curve. While the ACA and SMA probe the inner envelope ($\sim$1500\,au) response to accretion variability  where densities and temperatures are higher, the much larger beam of the JCMT includes a significant amount of colder and less dense envelope material ($\sim$6000\,au). This material is additionally subject to heating of the outer envelope by the external interstellar radiation field, and is located at much larger light travel times from the central protostar, which may have a significant effect on the shape of the JCMT light curve. 

In this section, we present our toy model for the EC 53 envelope structure and periodic outburst to explain the observed time delay between the SMA and JCMT. We then explore how varying the parameters of the toy model affects the observed light curves for EC 53. 

\label{sec:toy_model}

\subsection{EC 53 Model Description}
We consider a protostar with a varying luminosity surrounded by a spherically symmetric envelope with a fixed density structure. Based on the previous radiative transfer modeling of EC 53 by \cite{Baek2020}, the envelope is assumed to have a radial power law density profile,
\begin{equation}
\rho(r) = \rho_0 (r/r_0)^\alpha,
\label{eqn:density}
\end{equation}
truncated at an inner and outer radius, $r_\mathrm{in}$ and $r_\mathrm{out}$. The dust temperature profile follows a radial power law with a floor temperature below which the envelope can not fall, mimicking the effects of external heating by the (local) interstellar radiation field:
\begin{equation}
    T(t,r) = 
\begin{cases}
    T_0(t)\left[r / r_0\right]^{-2/(4 + \beta_\mathrm{em})},& \text{if } T\geq T_\mathrm{floor}\\
    T_\mathrm{floor},              & \text{otherwise}
\end{cases}.
\label{eqn:dust_temp}
\end{equation}
The power law portion of this temperature profile implicitly assumes the dust opacity also follows a power law with frequency, $\kappa_\nu \propto \nu^{\beta_\mathrm{em}}$, over the frequencies $\nu$ responsible for the bulk of the dust emission. Following \citet[][Section 6.2]{ContrerasPena2020}, the dust temperature profile in the envelope is related to the steady state luminosity by
\begin{equation}
 T_0(t) = C\left(\frac{L(t)}{L_\odot}\right)^{(1/(4 + \beta_\mathrm{em}))},
 \label{eqn:T_0_steady}
\end{equation}
where $L(t)$ is the time dependent accretion luminosity of the source and $C$ is a conversion constant which depends on the assumed dust properties and the reference radius $r_0$.

When the accretion luminosity of the protostar changes, the time required for the light pulse to travel outward through the envelope and heat the dust will result in a lag in the observed dust temperature response. \cite{Johnstone2013} performed analytic radiative transfer calculations of the propagation of accretion bursts through protostellar envelopes, and found the light travel time to be significantly longer than the dust heating timescale. We therefore assume that changes in the dust temperature  propagate 
at the speed of light. Photons emitted by the heated dust will then reach the observer with various lags, depending on their place of origin in the envelope.

To simplify our modeling of the envelope, we use a cylindrical coordinate system with azimuthal symmetry, where the positive $x$ direction points towards the observer, the positive $z$ direction is North on the sky, and the distance from the protostar is $r=\sqrt{x^2+z^2}$. We then calculate a lookback time for any position in the envelope as a combination of the time for light to propagate from the protostar to that location, $r/c$, minus the travel time from the relative offset of the position along the line of sight, $x/c$. Combined,
\begin{equation}
t_\mathrm{lb} = (r - x)/c,
\label{eqn:tlb}
\end{equation}
where $t_\mathrm{lb}=0$ is the time at which photons from a burst first reach the observer. During the burst, the observer sees a given position in the envelope to be experiencing a luminosity from the protostar of $L(t-t_\mathrm{lb})$, which modulates the dust temperature response, equation \ref{eqn:T_0_steady}, such that 
\begin{equation}
 T_0(t) = C \left(\frac{L(t-t_\mathrm{lb})}{L_\odot}\right)^{(1/(4 + \beta_\mathrm{em}))}.
 \label{eqn:T_0_mod}
\end{equation}

\begin{figure}[htb]
    \centering
    \includegraphics{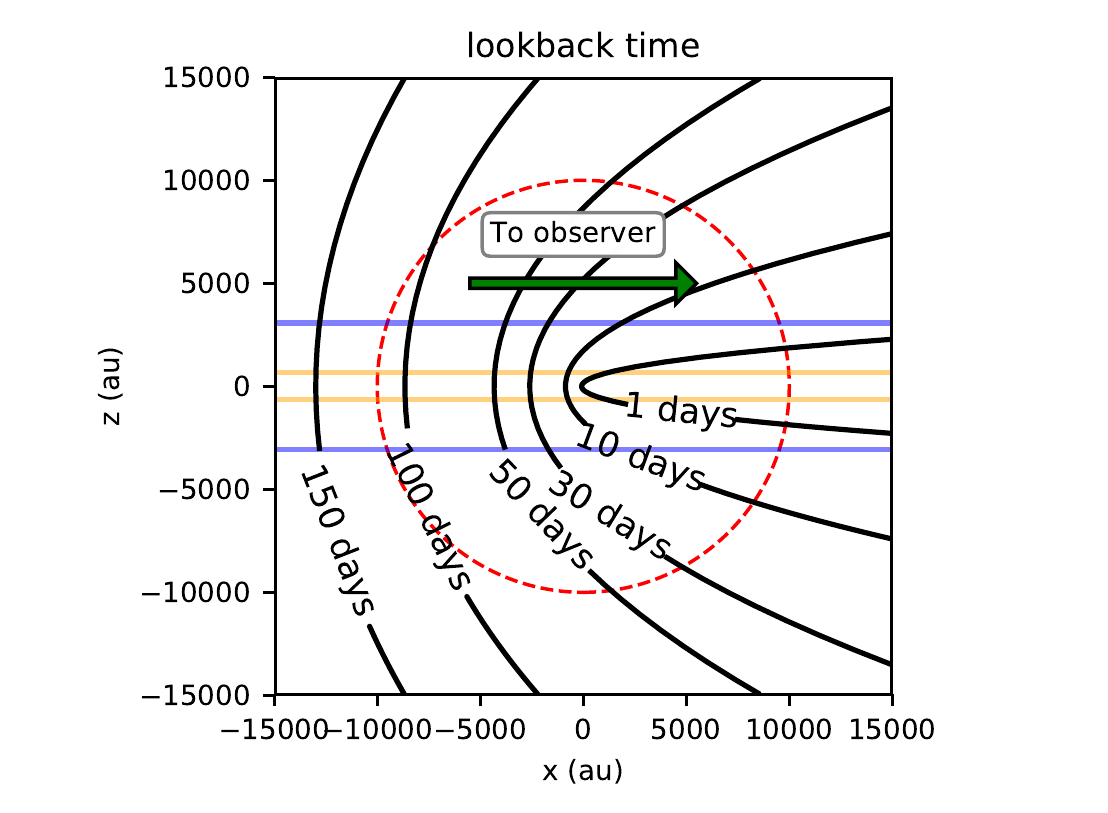}
    \caption{Lookback time contours over the domain used for our toy model envelope. The dashed red circle denotes the outer envelope edge, while the solid blue and orange lines show the radii of our model beams for the JCMT 850 $\mu$m (3270 au) and SMA (654 au).}
    \label{fig:lookback_time}
\end{figure}

To illustrate how the lookback time affects the observed light curve, in Figure \ref{fig:lookback_time} we show lookback time contours in days over the $30000 \times 30000$ au spatial domain used for our models, where a typical envelope of radius of $10000$ au is denoted by a dashed red circle, and the representative sizes of the SMA and JCMT (850 $\mu$m) beams at the distance of Serpens are overlaid. 

Regardless of the beam size, the timescale for the observer to see the entire spherical envelope respond to an instantaneous burst is simply the light travel time across the envelope diameter, $\approx 115$ days. However, the beam size of the telescope will modulate the observed response of the envelope: a larger beam will incorporate more envelope material with larger lookback times $t_\mathrm{lb}$, producing a more pronounced lag in the response relative to a smaller beam size.

To compute a model light curve from a given luminosity function, we discretize our model over a 2D $xz$-grid, where $x_j$ and $z_k$ denote the coordinates at the center of each cell, and sample the luminosity at timesteps $t_i$ with interval $\Delta_t$. The grid cells have a constant width of $\Delta_\mathrm{cell}$ in units of au, while the grid is $n_\mathrm{cell}$ and $n_\mathrm{cell}/2$ cells in the $x$ and $z$ directions respectively, owing to the symmetry about the $z=0$ axis. For each timestep, we calculate  the density and dust temperature using equations \ref{eqn:density}, \ref{eqn:dust_temp}, and \ref{eqn:T_0_mod}. The envelope is assumed to be optically thin throughout, such that the dust in each model cell emits as a blackbody with intensity
\begin{equation}
I_\nu(t_i,r_{j,k}) = \kappa_\nu \rho(r_{j,k}) B_\nu(\nu, T(t_i,r_{j,k})),  
\label{eqn:model_intensity}
\end{equation}
where $r_{j,k} = \sqrt{x_j^2 +z_k^2}$,
$B_\nu$ is the Planck function and $\nu$ is the observing frequency. 
Given that we are only interested in relative variations in the brightness of the protostellar envelope, fixed quantities in equation \ref{eqn:model_intensity}, such as $\kappa_\nu$, can be ignored or set to unity.

The flux is then calculated for a given telescope and instrument by integrating the model intensity in each cell within a representative circular beam:
\begin{equation}
F(t_i) = \sum_{j=0}^{n_\mathrm{cell}} \sum_{k=0}^{k_\mathrm{beam}}  2\pi z_k  I_\nu(t_i,r_{j,k})\Delta_\mathrm{cell}^2,
\end{equation}
where $k_\mathrm{beam} = \lfloor \frac{\theta_\mathrm{beam}/2}{ \Delta_\mathrm{cell}} \rfloor$ and $\theta_\mathrm{beam}$ is the beam FWHM in au. The representative observing frequencies and beam sizes used for our models are listed in Table \ref{tab:model_telescopes}.

\begin{deluxetable}{cccc}
\tablecaption{Toy Model Observing Properties}
\label{tab:model_telescopes}
\tablehead{\colhead{Observatory} & \colhead{Beam size (\arcsec)} & \colhead{Beam radius (au)} & \colhead{Frequency (GHz)}} 
\startdata
JCMT 850 \micron & 15 & 3270 & 352.94 \\ 
JCMT 450 \micron & 10  & 2180 & 666.66 \\
SMA  1300 \micron & 3  & 654  & 230.00 \\
ACA   850 \micron   & 4  & 872  & 343.50 
\enddata
\tablecomments{Beam size is the full width at half maximum. The beam radius is half this value converted to au assuming the model source lies at the distance to Serpens Main of 436 pc \citep{Ortiz-Leon2017}.}
\end{deluxetable}

\subsection{Fiducial EC53 Envelope Model}

The parameters for our fiducial EC 53 model are based on the best-fit 2D radiative transfer models by \citetalias{Baek2020}, with some simplifications. \citetalias{Baek2020} fit the structure of EC 53 using the observed SED during the quiescent phase, and altered the luminosity of the central protostar to reproduce the observed brightening at 850 $\mu$m in the JCMT Transient Survey. We use the same density structure for the envelope as \citetalias{Baek2020} (equation \ref{eqn:density}), with an $r_\mathrm{out}$ of 10000 au and $\alpha=-1.5$. While \citetalias{Baek2020} also included a 90~au circumstellar disk and $20^{\circ}$ opening-angle outflow cavity in their models, we do not include either, as these features are primarily important for variability in the near- to mid-infrared emission and should only have a modest effect on the broad shape and time delay seen in the submm emission at the angular scales probed by the JCMT and ACA. Thus, in this paper we only consider spherically symmetric density profiles, leaving more complex geometries to a future work. 

To convert our model luminosities to a dust temperature profile, we set a value for the constant $C$ in equation \ref{eqn:T_0_steady}, which implicitly depends on the details of the dust opacity. We derive a value $C$ = 63 K $(100~\mathrm{au})$ from a by-eye fit of the temperature profiles for the \citetalias{Baek2020} radiative transfer models in the outbursting and quiescent phases of EC 53 (Fig 10 of \citetalias{Baek2020}) using equations \ref{eqn:dust_temp} and \ref{eqn:T_0_steady}.

For the time-dependent luminosity, we follow the brightness model of EC53 used by \citet[their equation 23]{Lee2020yh}  to interpret the phase-folded mid-infrared and submm light curves of EC 53. This luminosity function uses a periodic exponential rise and decay and is motivated by the shape of the near infrared light curves, which should approximately trace changes in the protostellar luminosity. Thus, our luminosity function is:
\begin{equation}
    L(t) = 
    L_\mathrm{max}\left(       e^{-t_\mathrm{mod} / \tau_\mathrm{fall}} +
   f  e^{(t_\mathrm{mod} - P) / \tau_\mathrm{rise}} \right),
   \label{eqn:EC53_lum}
\end{equation}
where $P$ is the period, $t_\mathrm{mod} = t\pmod P$, and $f = (1 - e^{-P / \tau_\mathrm{fall}})$ is a scaling factor which ensures a continuous transition between the rise and fall parts of the function. We set the rise and fall timescales to the values determined empirically by \cite{Lee2020yh} from fits to the near infrared light curves of $\tau_\mathrm{rise} = 35$ days and $\tau_\mathrm{fall} = 270$ days. The period and maximum luminosity are allowed to vary freely however, as the exact duration between past outbursts of EC 53 and their peak brightness exhibit some variability \citep{Lee2020yh}.

To compare the effect of our model light curve against the submm observations, we compute the model flux for each telescope in Table \ref{tab:model_telescopes} over a grid with $\Delta_\mathrm{cell} = 10$ au and $n_\mathrm{cell} = 3000$, and sample the luminosity function at timesteps of length $\Delta_t=0.5$ days. Since our model does not include the details of the dust mass and opacity needed to output the brightness in physical units, we normalize the model light curves to their minima.

\begin{figure}[htb]
    \centering
    \includegraphics[]{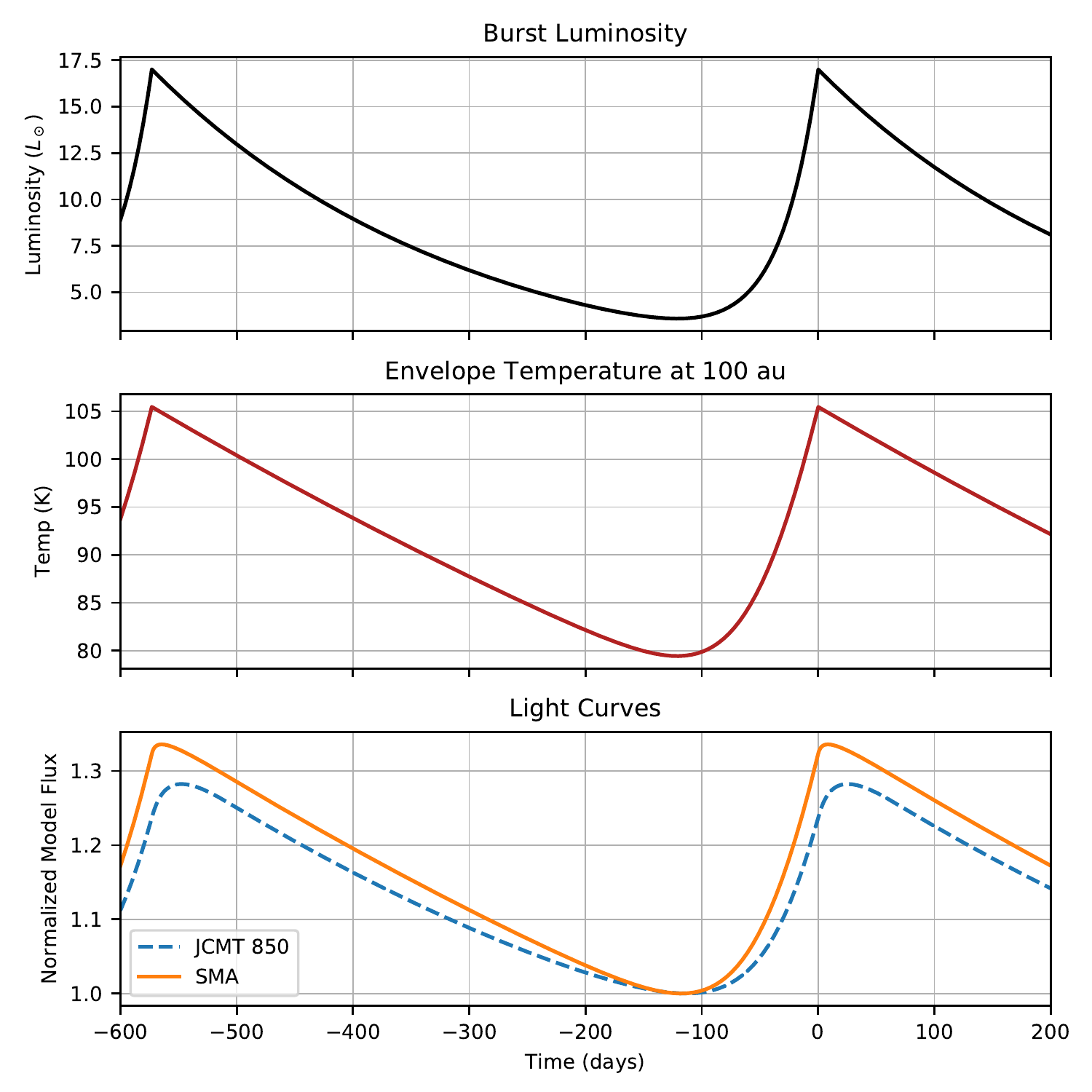}
    \caption{Fiducial EC 53 model luminosity function (top panel), envelope dust temperature at 100 au (center panel), and normalized model SMA and JCMT 850 $\mu$m light curves (bottom panel). The parameters of the model are presented in Table \ref{tab:ec53_fiducial}.}
    \label{fig:ec_53_fid_fiducial_model}
\end{figure}

\begin{figure}[htb]
    \centering
    \includegraphics[]{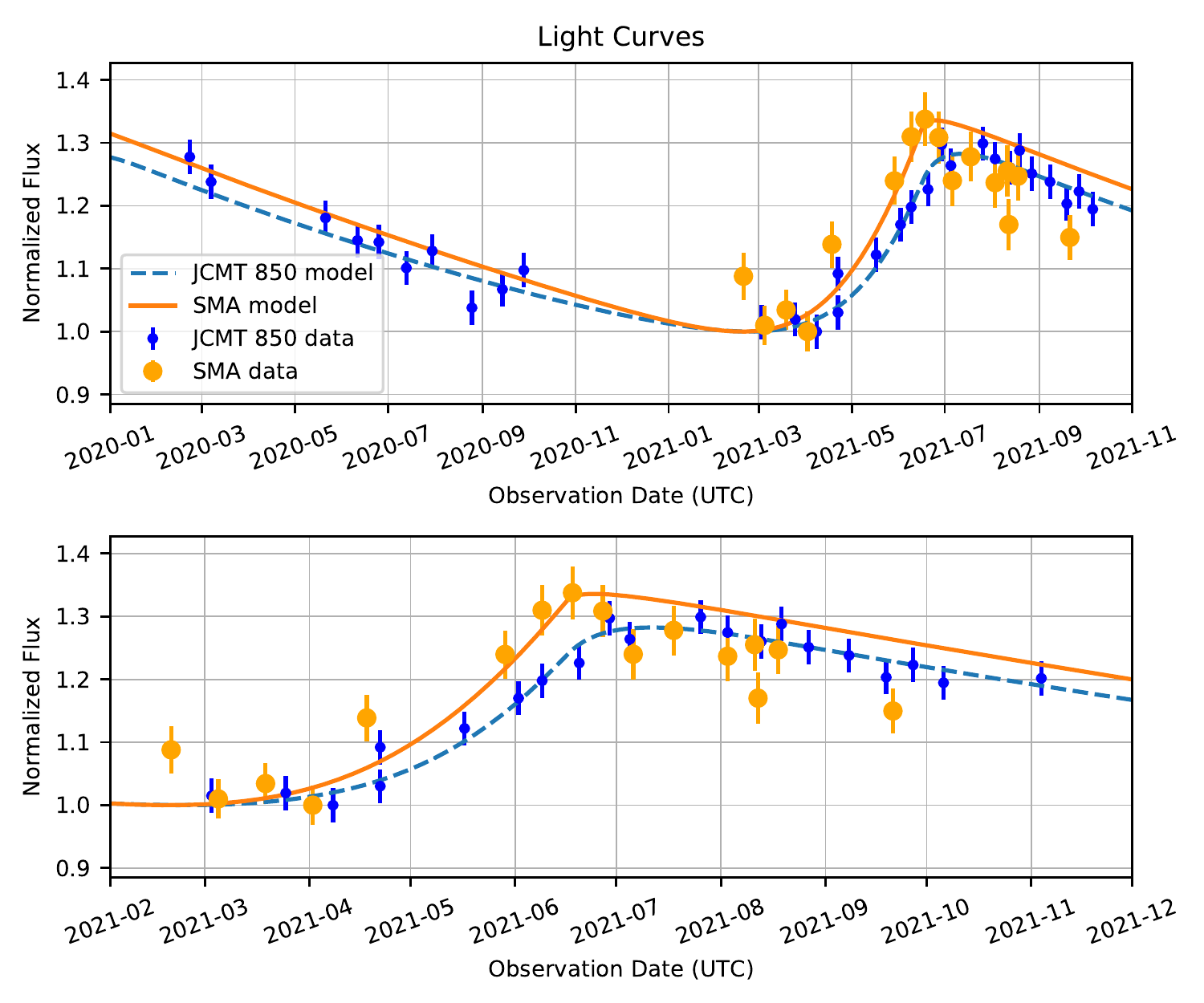}
    \caption{Comparison of our fiducial EC 53 model light curves and the observations for the SMA and JCMT (850 $\mu$m). The models and observations are shown over the 2020 decay and 2021 outburst (top panel) and zoomed in on the 2021 outburst (bottom panel). The model light curves are normalized to the minimum flux, while the data are normalized as in the lower panel of Figure \ref{fig:sma_lc}.}
    \label{fig:ec_53_fid_model_fit}
\end{figure}

To fit the observations, we vary the date of the peak luminosity ($t=0$ in equation \ref{eqn:EC53_lum}), period, floor temperature, and maximum luminosity of the model to match the observed light curves by eye. We find good agreement using $t_\mathrm{peak} = $ 2021-06-16, $P=573$ days, $T_\mathrm{floor} = 24$ K, and $L_\mathrm{max} = 17 L_\odot$; the corresponding minimum luminosity is $L_\mathrm{min} = 3.6 L_\odot$ and thus $L_\mathrm{max}/L_\mathrm{min} = 4.7$.  The parameters of our fiducial model are summarized in Table \ref{tab:ec53_fiducial}, the model luminosity, temperature at $r_0=100$ au, and SMA/JCMT (850 $\mu$m) light curves are shown in Figure \ref{fig:ec_53_fid_fiducial_model}, and comparison of the observed SMA/JCMT and model light curves is given in Figure \ref{fig:ec_53_fid_model_fit}. We do not perform formal model fitting, as our purpose is only to qualitatively reproduce the relative amplitude and time delay of the EC 53 outburst rather than constrain exact parameter values. We have also computed model light curves for the ACA and the JCMT 450 $\mu$m observations, however, comparison with the EC 53 observations is more complicated owing to the paucity of data during the 2021 outburst and additional observational uncertainties; we thus defer our discussion of these models to Section \ref{sec:disc}. 

Our model light curve is in good agreement with the JCMT 850 $\mu$m observations for both the most recent decay in 2020 and the subsequent outburst in 2021.  We also reproduce the relative amplitude and earlier rise of the SMA observations with respect to the JCMT, providing strong evidence for modulation of the accretion burst during its propagation through the envelope. Our fiducial model underestimates the rate of brightness decline at the SMA after the burst, however, suggesting that some details of the luminosity function and/or envelope properties are missing, which we explore further in Section \ref{sec:further_modeling}.

\begin{deluxetable*}{ll}
\tablecaption{EC 53 Fiducial Model}
\label{tab:ec53_fiducial}
\tablehead{\colhead{Parameter} &  \colhead{Value}} 
\startdata
Model Grid & \\
\hline
$n_\mathrm{cell}$ & 3000 \\
$\Delta_\mathrm{cell}$ & 10 au \\
\hline
Envelope Properties & \\
\hline
$r_\mathrm{in}$        & 0 au \\
$r_\mathrm{out}$       & 10000 au \\
$r_\mathrm{0}$         & 100 au \\
$\rho_0$               & 1.0 (arb. units) \\
$\alpha$               & -1.5 \\
$\beta_\mathrm{em}$    & 1.5 \\
$T_\mathrm{floor}$     & 24 K \\
$C$                    & 63 K $(r_0=100~\mathrm{au}$) \\
\hline
Burst Properties and Time Sampling & \\
\hline
$L_\mathrm{max}$      & 17 $L_\odot$ \\
$L_\mathrm{min}\tablenotemark{a}$      & 3.6 $L_\odot$ \\  
$\tau_\mathrm{rise}$  &  33 days \\ 
$\tau_\mathrm{fall}$  &  270 days \\ 
$P$    & 573 days \\
$\Delta_t$ & 0.5 days \\ 
$t_\mathrm{peak}$ & 2021-06-16 (UTC)
\enddata
\tablecomments{\tablenotetext{a}{The minimum luminosity is dependent on the other parameters of the luminosity function, see equation \ref{eqn:EC53_lum}.}
}
\end{deluxetable*}

\subsection{EC 53 Fiducial Model Parameter Space Exploration}
In this Section, we explore how variations in selected parameters of the fiducial EC 53 model affect the relative amplitude and time lag of the observed light curves to evaluate the robustness of the fiducial model and illustrate the role each parameter plays.

\subsubsection{Varying EC 53 Envelope Properties}

\begin{figure}[htb]
    \centering
    \includegraphics[width=0.8\textwidth]{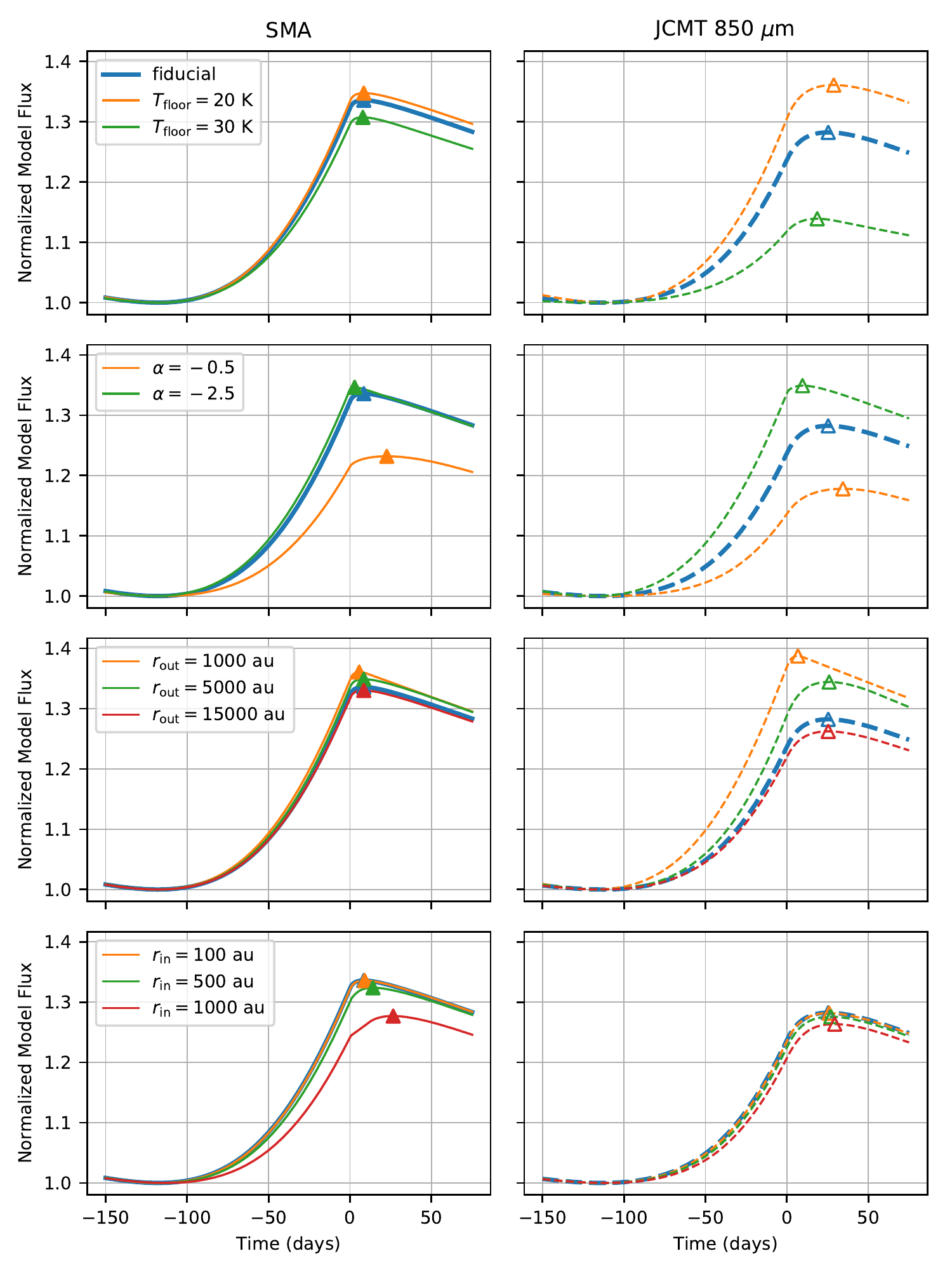}
    \caption{Comparisons of our EC 53 model light curves for the SMA (left column) and JCMT (right column) with variations in the envelope parameters. The fiducial model is marked in each panel by the blue line and the origin on the time denotes the peak brightness of the source (see Figure \ref{fig:ec_53_fid_fiducial_model}). The triangles show the peak position for each curve. The varied parameters are as follows: \textit{first row:} envelope floor temperature $T_\mathrm{floor}$; \textit{second row:} envelope density power law index $\alpha$; \textit{third row:} envelope outer radius $r_\mathrm{out}$; \textit{fourth row:} envelope inner radius $r_\mathrm{in}$.}
    \label{fig:envelope_comp}
\end{figure}

In Figure \ref{fig:envelope_comp} (top row), we show the effect on the model SMA and JCMT (850 $\mu$m) light curves of varying the envelope floor temperature $T_\mathrm{floor}$, envelope density radial index $\alpha$, outer envelope radius $r_\mathrm{out}$, and inner envelope radius $r_\mathrm{in}$ with the maximum amplitude in each model light curve marked by a triangle. The floor temperature in the model is most important for the JCMT response, as the radius in the fiducial model where $T_\mathrm{floor}=24$ K is reached ($\sim 2700$ au at burst minimum) is fully enclosed within the JCMT beam. With a higher floor temperature of $T_\mathrm{floor}= 30 $ K, this radius is significantly smaller ($\sim 1500$ au), resulting in less of a temperature response from the outer envelope, and therefore significantly reducing the JCMT burst amplitude. A weaker reduction in amplitude is seen for the SMA, as the material near the floor temperature is only located along the front and back of the beam column, rather than the sides as in the case of the JCMT. With a lower floor temperature of $T_\mathrm{floor}= 20$ K the JCMT burst amplitude is significantly increased, as the radius at which $T_\mathrm{floor}$ is reached ($\sim 4300$ au) now occurs well outside the JCMT beam. The time delay between the SMA and JCMT light curves slightly increases with a lower $T_\mathrm{floor}$, due to the increase in the light travel time to radii where the floor temperature is important.  

We next vary the radial density power-law index (Figure \ref{fig:envelope_comp}, second row from the top). We note that the mass within the envelope is not conserved when varying the density index, however, our normalization of the light curves to the minimum flux removes the differences in envelope luminosity resulting from the differing masses. With a steeper density profile, $\alpha=-2.5$, little change from the fiducial model is seen in the SMA light curve. The amplitude of the JCMT light curve increases, however, and there is less delay between its peak and that of the SMA. The modified JCMT light curve is the result of concentrating more of the envelope within the center of the JCMT beam, reducing the influence of cold outer envelope material with temperatures close to $T_\mathrm{floor}=24$ K and longer lookback times. When the density profile is shallower, $\alpha=-0.5$, the peak amplitude for both the SMA and JCMT light curves is significantly reduced and the delay time between the peaks is slightly shorter. The reduction in normalized amplitude occurs because there is a significant increase in the amount of colder dust at large radii, which has a weak brightness response due to the temperature floor. For both the JCMT and ACA, the lag between the protostellar burst and the observed peak increases as the outer envelope becomes more important. This lag is more pronounced for the SMA, resulting in a shortening of the relative lag between the JCMT and the SMA.

The effect of changes to the envelope outer radius $r_\mathrm{out}$ (Figure \ref{fig:envelope_comp}, 2nd row from bottom) depends on its size relative to the model beams (refer to Table \ref{tab:model_telescopes}). For $r_\mathrm{out}= 15000$ au, very little change from the fiducial is seen other than a slight amplitude reduction, which is the result of adding only a small amount of cold material to the front and back of the beam columns along the observer's line of sight. For a smaller than fiducial envelope ($r_\mathrm{out} = 5000$ au), the SMA amplitude only slightly increases, while the JCMT amplitude increases significantly due to the removal of cold material close to the temperature floor. In the extreme case of a very small, $r_\mathrm{out} = 1000$  au envelope, which is fully enclosed within the JCMT beam, the shape of the JCMT and SMA light curves become nearly identical and there is a negligible time delay. The JCMT reaches a higher peak amplitude due to the removal of all envelope material at large radii where the lookback times are long and the temperature approaches the floor.

While the envelope in our fiducial model extends all the way to the grid center, we experiment with truncating the envelope some distance from the central protostar to determine if envelope substructure, e.g., low density central cavities carved by disks, may have an identifiable effect on the light curve (Figure \ref{fig:envelope_comp}, bottom row). With an inner cavity of radius $r_\mathrm{in} = 100$ au, the light curves for both the SMA and JCMT are unchanged. Despite the radial power law density and temperature structure of the envelope, the submm brightness within 100 au is negligible in comparison with the contribution from the envelope material out to $r=10000$ au along the observer's line of sight and within either beam. For an $r_\mathrm{in} = 500$ au cavity, the removed material becomes more significant and slightly reduces the SMA peak amplitude and shifts forward the peak date, while the JCMT model light curve is still unchanged. If the cavity is sufficiently large to be resolved by the SMA beam ($r_\mathrm{in} = 1000$ au), a large reduction in the SMA peak amplitude occurs, and a ``kink'' appears in the light curve at $t=0$. The large hole effectively disconnects the emission from the near and far sides into two pulses separated in time, with the near-side modulated by a narrow, 10 day spread, while the far-side begins later and has a much longer response spread (see Figure \ref{fig:lookback_time}). At the scale of 1000 au, the cavity is still unresolved by the JCMT beam, but a slight reduction in peak amplitude occurs. This highlights the possibility for envelope substructure to modulate the response to an accretion burst in an observable way, provided sufficiently high resolution monitoring. In reality, outflow cavities are likely significantly more complex than the simple spherical holes we have used here, however more accurate 3D modeling is beyond the scope of this paper.

\subsubsection{Varying EC 53 Burst Properties}
\label{ssec:ec53_burst_prop}
\begin{figure}[htb]
    \centering
    \includegraphics[width=0.8\textwidth]{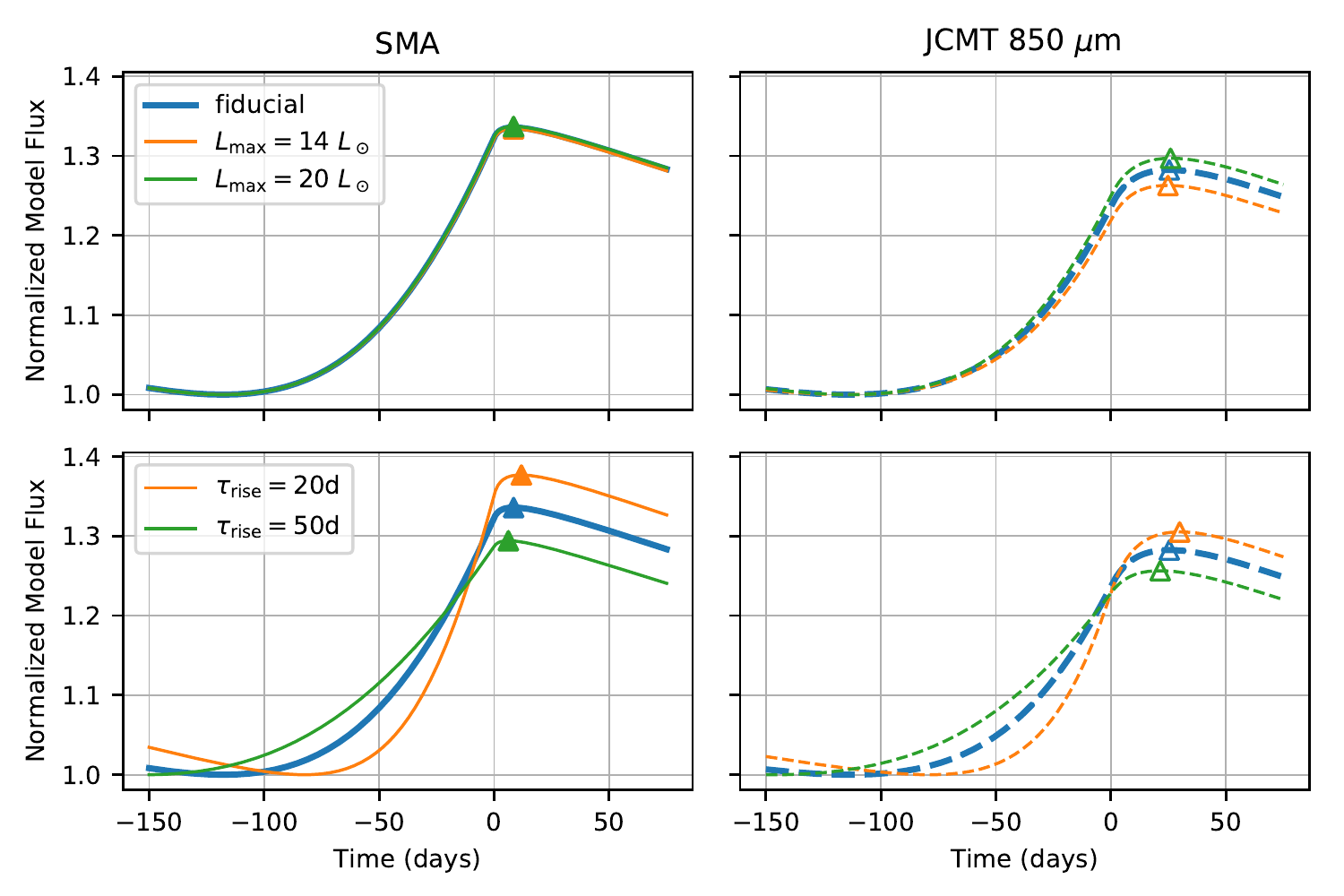}
    \caption{As Figure \ref{fig:envelope_comp}, but for comparison of the luminosity function parameters, see equation \ref{eqn:EC53_lum}. \textit{Top row:} maximum luminosity $L_\mathrm{max}$; \textit{Bottom row:} rise timescale $\tau_\mathrm{rise}$.}
    \label{fig:burst_comp}
\end{figure}

In Figure \ref{fig:burst_comp} we consider the effect on the model SMA and JCMT (850 $\mu$m) light curves of changing the maximum luminosity, $L_\mathrm{max}$, and the rise timescale, $\tau_\mathrm{rise}$, of our fiducial luminosity function. Modifying the maximum luminosity has essentially no effect on the amplitude of the outburst seen by the SMA, as the minimum luminosity reached in equation  \ref{eqn:EC53_lum} remains a fixed fraction of the peak luminosity and the majority of the emission in the SMA beam is from material radiating at high enough temperatures (see Equation \ref{eqn:T_0_mod}) to always be in the Rayleigh-Jeans limit. The amplitude of the burst at the JCMT is slightly increased, as more of the outer envelope stays above the floor temperature of 24 K, and thus undergoes a greater temperature change during the outburst. The delay time between the SMA and JCMT peak remains unchanged. 

The rise timescale affects the slope of the light curve and burst amplitude for both the SMA and JCMT. A shorter rise time increases the burst amplitude, as the decay from the maximum luminosity will continue for a longer time before the exponential rise takes over, and the minimum luminosity --- at which our light curves are normalized --- will be lower. Similar changes to the burst amplitude occur if the period and fall timescales are modified, but are less dramatic owing to their much longer lengths than the rise timescale. While the burst peaks earlier at both the SMA and JCMT with a shorter rise time, the time delay between the SMA and JCMT is not strongly affected.

\section{Observing Generic Accretion Bursts}
\label{sec:further_modeling}

In this section, motivated by the differences in outburst behaviour seen between the SMA, ACA, and JCMT light curves of our variable targets, we explore how different types of accretion bursts propagate through our fiducial EC 53 envelope model (Table \ref{tab:ec53_fiducial}). To survey the prospects for monitoring protostellar variability with other telescopes, we also experiment with how the telescope beam size and observing wavelength determines the burst response.

\subsection{Modulation of Short Bursts, Brightness Jumps, and Periodic Variability}
\label{ssec:washouts}

Our fiducial model, based on the best-fit 2D radiative transfer analysis by \citetalias{Baek2020}, works very well in the mean. However, after the peak, the SMA decay time for EC 53 is less steep than observed (Figure \ref{fig:ec_53_fid_model_fit}). Near infrared observations of EC 53 with a high cadence have shown structure in the light curve during the peak \citep[][Figure 5]{Lee2020yh}, suggesting that there is additional behaviour in the luminosity variability that our simple model (equation \ref{eqn:EC53_lum}) does not capture. Similarly, in the ACA monitoring of SMM 10, strong stochastic variability is found, whereas the contemporaneous JCMT light curves show only moderate variability. For both sources, these empirical differences could be the result of a poorer sensitivity to rapid variations in the envelope brightness obtained with the larger beam of the JCMT, $\sim15\arcsec$, compared against the $\sim3-4\arcsec$ beams of the SMA and ACA. 

\begin{figure}[htb]
    \centering
    \includegraphics{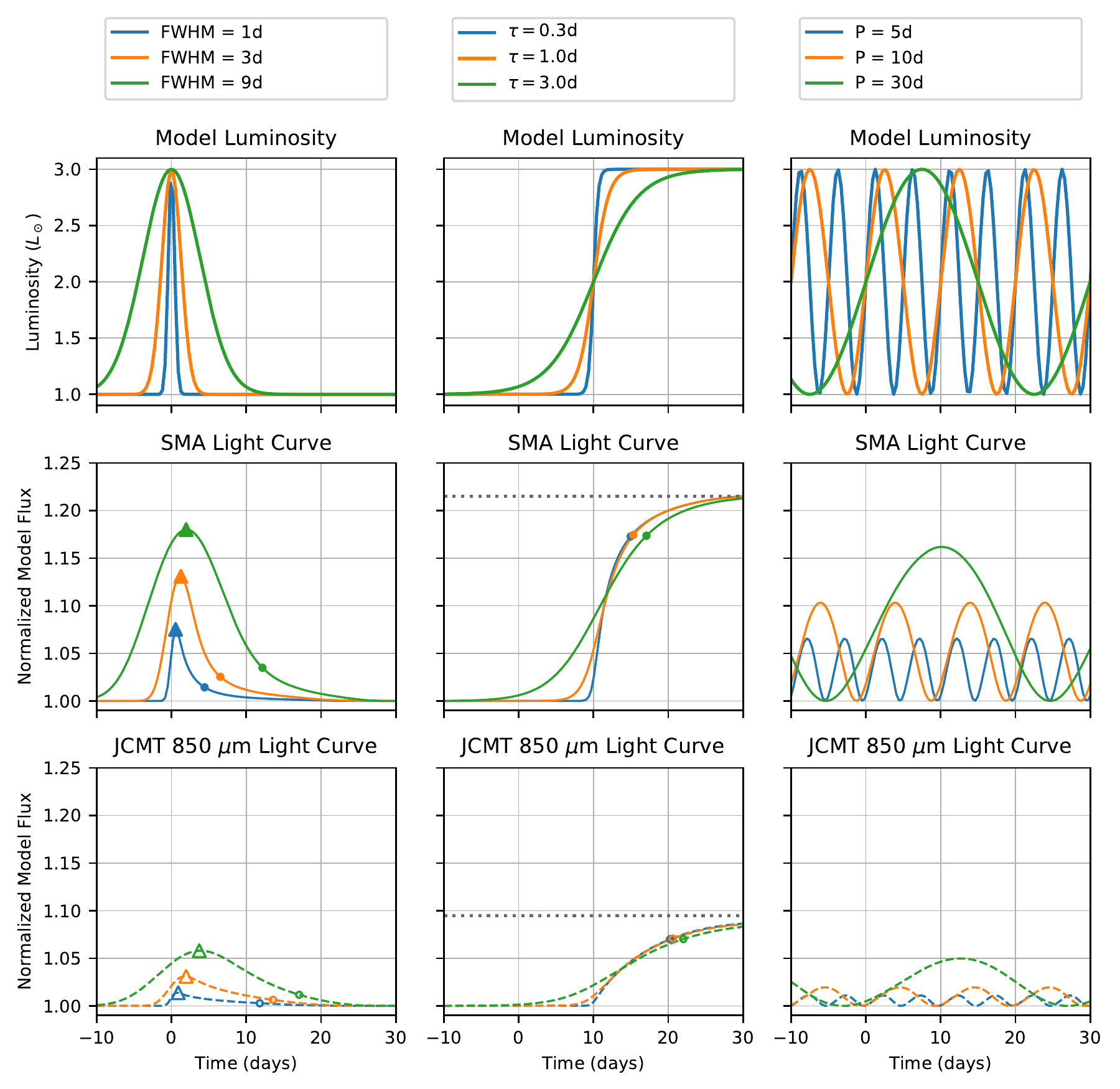}
    \caption{Comparison of the response of the fiducial EC 53 envelope model to Gaussian (left column), sigmoid (center column), and sin wave (right column) luminosity functions with various shapes. \textit{Top row:} input luminosity function; \textit{middle row:} normalized model SMA light curve; \textit{bottom row:} normalized model JCMT light curve. For the Gaussian model, the maximum flux and time during the decay when 20\% of the maximum flux is reached are shown by triangles and circles respectively. For the sigmoid model, circles mark the time when 80\% of the maximum flux is reached, which is shown by the dotted line. 
    Max=triangles, 20\% max=circles.}
    \label{fig:gauss_sin_comp}
\end{figure}

We demonstrate the variation in observed response at the SMA and JCMT to various types of luminosity changes in Figure \ref{fig:gauss_sin_comp}, using the fiducial EC 53 envelope model. Here we compare fixed amplitude Gaussian, sigmoid, and sine wave luminosity functions (top row) with varying FWHM (1--9 days), rise timescale $\tau$ (0.3 -- 3 days, where $L \propto 1/(1+e^{-t/\tau})$) or period (5--30 days), respectively, against the resulting model light curves for the SMA (middle row) and JCMT (bottom row).  For the case of a Gaussian outburst, we mark the time at which the model brightness peaks and the time during the decay when 20\% of the maximum is reached, for both telescopes. Similarly, for the sigmoid outburst we mark the time at which the model brightness reaches 80\% of the maximum. 

The Gaussian outbursts of shorter duration yield lower observed light curve amplitudes for both the SMA and JCMT, however, in all cases the observed amplitude is a factor of $\sim 2$ smaller for the JCMT. Furthermore, the lag timescale for the flux to return to the 20\% level is significantly longer, 17--21 days for the JCMT versus 4--9 days for the SMA. 

For the sigmoid outbursts, we see a similar difference in amplitude between the SMA and JCMT light curves as the Gaussian case, and for all values of $\tau$, the SMA model brightness reaches the 80\% level sooner than the JCMT (15--18 vs 20-22 days). For both SMA and JCMT observations, differences in the rise timescale are easier to detect early in the burst than later, as the light curve response at early times is dominated by large changes in brightness in the inner envelope where the lookback time is shorter. Conversely, after the 50\% amplitude level is reached, the differences in the rise timescale $\tau$ are harder to detect, as the response is diluted by envelope material at larger radii with long lookback times. 

With a sine wave luminosity function the amplitude of the model light curves are similarly reduced for shorter periods. The longer lag time of the JCMT response, however, results in an additional washing out of the light curve amplitude. Thus, it is entirely plausible that rapid, $< 30$ day, variations in protostellar luminosity may be responsible for both the observed structure in the post-peak decay of the SMA EC 53 light curve (Figure \ref{fig:sma_lc}), and the factor of $\sim 3$ stronger stochastic variability observed between the ACA and JCMT (850 $\mu$m) light curves of SMM 10 (Figure \ref{fig:aca_lc}, bottom panel).

\subsection{Effect of Beam Size and Observing Wavelength}
\label{ssec:beam_size_and_nu_comp}

\begin{figure}[htb]
    \centering
    \includegraphics[]{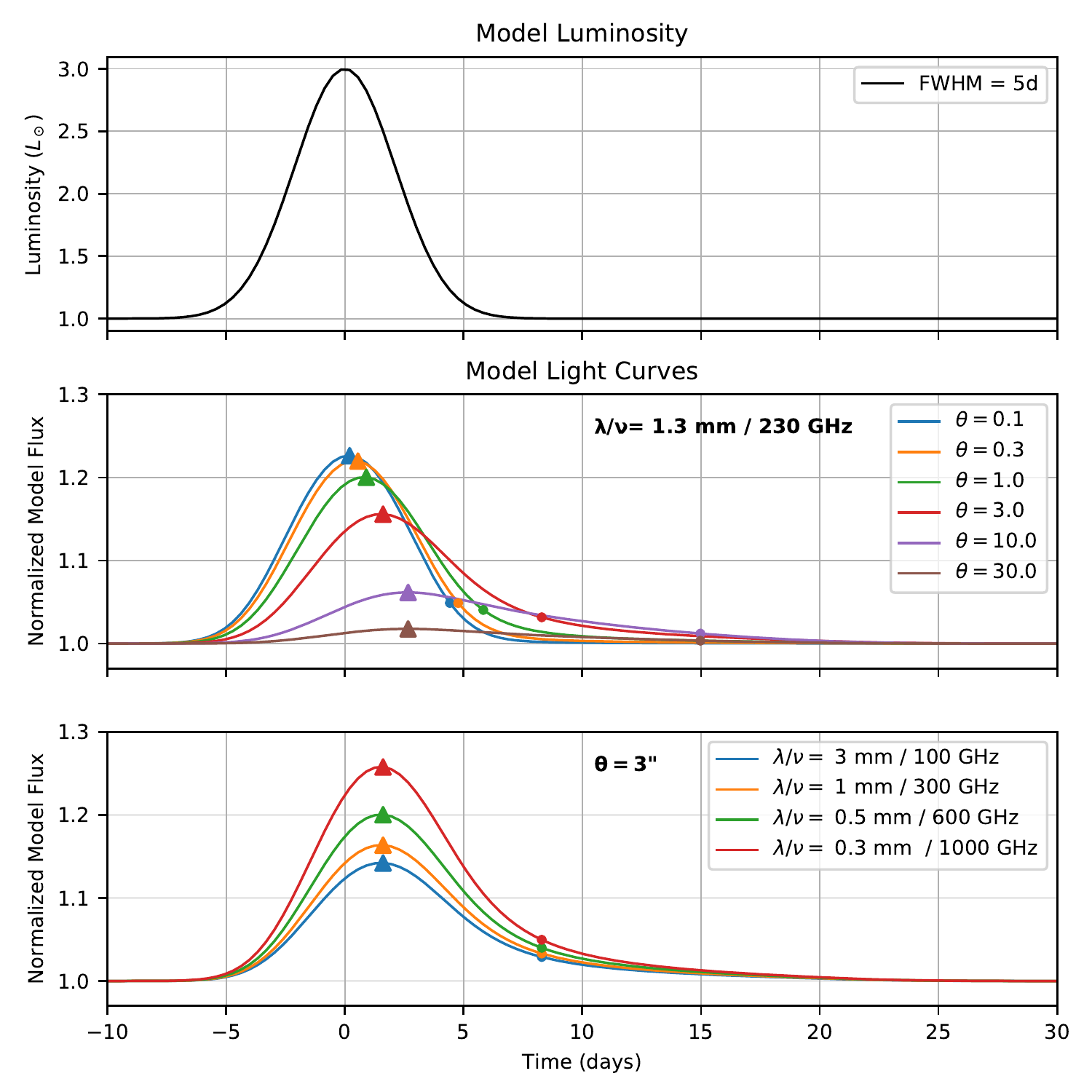}
    \caption{As Figure \ref{fig:gauss_sin_comp}, but for comparison of a different observing properties with a Gaussian luminosity function. \textit{Top row:} Luminosity function profile; \textit{middle row:} normalized light curve for a fixed wavelength and varying beam size $\theta$; \textit{bottom row}: normalized light curve for a fixed beam size and varing wavelength.}
    \label{fig:obs_comp}
\end{figure}

The observed envelope response to a change in protostellar luminosity is sensitive to both the telescope beam size and observing wavelength, which differ between our ACA, SMA, and JCMT measurements. To illustrate the effect of these properties on the observations, in Figure \ref{fig:obs_comp} we show model light curves for a Gaussian luminosity function (FWHM = 5 days, top panel) with varying beam size (center panel) and observing wavelength (bottom panel). The range of beam sizes correspond to high resolution observations with an interferometer, such as the ALMA 12m array ($0.1\arcsec$), up to single dish observations with the envelope fully unresolved ($30 \arcsec$). For Serpens, this corresponds to physical sizes ranging from 40 to 12000\,au. With a beam size $\leq 1.0 \arcsec$, very little modulation of the luminosity function occurs, and the model light curves closely trace the underlying accretion luminosity variations. For increasing beam sizes, the time for the burst to propagate through a larger portion of the envelope introduces a significant reduction in amplitude ($< 50\%$) and an additional lag in the light curve. 

An decrease in observing wavelength has little effect on the light curve lag, but does produce a stronger amplitude for the response, e.g., the response in the far infrared (0.3 mm) is $1.6-1.8$ times stronger than typical mm observing wavelengths (1--3 mm). This increase in response amplitude is due to the fact that at shorter wavelengths a larger fraction of the envelope lies below the temperature for which the response is Rayleigh-Jeans (i.e.\ the brightness responds as $T^\alpha$, with $\alpha > 1$ rather than $\alpha \sim 1$; \citealt{Johnstone2013,ContrerasPena2020}).

\section{Discussion}
\label{sec:disc}

\subsection{Constraining Envelope Models and Accretion Burst Properties}
\label{ssec:general_disc}
Our work demonstrates that variations in the protostellar accretion rate produce a response in the dust temperature throughout the envelope, which yields a light curve modulated by the envelope structure and properties of the observations. Contemporaneous observations at multiple resolutions and wavelengths combined with modeling of the burst propagation through the envelope can thus recover detailed information about the envelope structure and identify the underlying accretion variations. Constraining the envelope structure used in modeling of the burst response is important for separating the accretion luminosity variations from their modulation by the envelope. The \citetalias{Baek2020} model for EC 53 used here as a basis for our toy model was determined through fitting of the observed SED with radiative transfer simulations; without such a template, forward modeling of the accretion luminosity variations would be much more uncertain (Section \ref{ssec:ec53_burst_prop}) due to the poorly constrained envelope properties.

High resolution observations are particularly important for identifying rapid variations (Section \ref{ssec:washouts}, Figure \ref{fig:gauss_sin_comp}); the response to a short burst with a large beam suffers from a longer lag time due to long lookback times, smaller amplitude owing to the inclusion of envelope material subject to non-variable heating by the ISRF, and potentially destructive interference in the response due to the potential propagation of multiple burst components within the beam. Rapid variations in the source brightness have a special physical interest, as they trace dynamical scales tied to the innermost disk where star-disk interface instabilities likely manifest \citep[e.g.][]{dangelo2012,armitage2016}.

Our ACA observations of SMM 10 (Figure \ref{fig:aca_lc}) and the near infrared light curve of EC 53 \citep{Lee2020yh} both show evidence for enhanced variations not witnessed in the lower resolution JCMT observations, suggesting that washing out of higher frequency variability is indeed occurring. Variability in the mid-infrared on timescales of days to months has been observed for many YSOs \citep[e.g.][]{Rebull2014, Park2021}, though a wide range of phenomena besides accretion variability can be responsible (e.g., rotating disk warps and starspots). Several notable examples show variability attributed to periodic accretion instabilities (L1634 IRS 7 \citep{Hodapp2015}, V347 Aurigae \citep{Dahm2020}, see also \citealt{Guo2022}), pulsed accretion from interaction with a binary (L1527 IRS \citep{Cook2019}), or rotation of a warped inner disk (LRLL 54361 \citep{Muzerolle2013}).

Investigating such sources with higher cadence and resolution submm/mm observations is important for identifying the nature and distribution of rapid accretion variability in protostars, but requires also high angular resolution.

High resolution submm/mm observations of the dust continuum can also provide valuable measurements of the inner envelope and protoplanetary disks dust density structure for modeling the burst propagation. Future observing campaigns can benefit from employing higher resolution, shorter wavelengths, and a variety of cadences. Long baseline observations with the ALMA 12m array would trace spatial scales associated with the dynamical timescales in the inner disk, and should have a stronger response to rapid variations in accretion luminosity (Section \ref{ssec:beam_size_and_nu_comp}). Furthermore, resolving the disk would allow its response to the burst to be separated from that of the envelope and therefore carefully quantified. 

Multi-wavelength monitoring can also provide complementary information on the physics of accretion variability. In particular, the typical dust temperatures in the envelope produce emission such that far infrared observations should produce a stronger light curve response, roughly proportional to the change in accretion luminosity, whereas submm/mm wavelength observations probe changes in brightness more closely tied to the induced dust temperature variations (see Section \ref{ssec:beam_size_and_nu_comp} and \citealt{Johnstone2013}). Near infrared observations should also trace protostar variability, but are subject to the effects of potentially time-varying extinction and emission from the hot inner disk \citep{Lee2020yh, hillenbrand2018, hillenbrand2019}. The combination of simultaneous observations at a variety of wavelengths should thus allow more accurate modeling of accretion luminosity variations and circumstellar environment; such an approach was used by \cite{Lee2020yh} to identify an increased buildup of material in the inner disk of EC 53 prior to outburst. Future planned ground-based observatories operating at the shortest submm wavelengths, such as CCAT-prime \citep{ccat2021}, and potential far infrared space-based missions \citep{andre2019, fischer2019} will be ideal for such extended monitoring.

\subsection{Observational Limitations and Model Caveats}
\label{ssec:caveats}

The observations and modeling of accretion variability presented in this paper have a variety of limitations, which we detail here. Although our relative flux calibration schemes provide an unprecedented level of accuracy ($\sim 3$\% vs 10-20\% with standard calibration schemes, see Section \ref{sec:rel_calibration}), they are still the dominant contribution to the error budget in our flux measurements, as all of our targets are mm bright and typically have an S/N $> 100$ per epoch. Observations of additional sources for relative calibration and relaxing the approximation of point-like calibrators may enable an even higher relative flux calibration accuracy. Our approach to flux measurement with the ACA and SMA observations is robust but simplistic, and only considers the point-like emission within the brightest source in the map. We assume the extended emission surrounding the variable targets to be stable and remove it during model fitting, however, in principle light echoes from the propagation of the accretion burst through the extend emission may be observable provided sufficient imaging fidelity. Such a measurement is challenging given the $uv$-coverage with the limited number of baselines at the ACA and SMA. Observations with improved $uv$-coverage using the 12m ALMA array and imaging techniques better suited to reconstruction of resolved sources than the standard tclean \citep[e.g.][]{Akiyama2017,Chael2018,Honma2014} may provide improved models of our sources and allow measurements of variability over the entire field of view. For the single-dish observations, better understanding of the JCMT beam is likely required before any significant further improvement in the relative flux calibration is possible, especially at 450 \micron\  \citep{Dempsey2013,mairs2021}.

Our toy model of EC 53 is very simplistic. It is based on the envelope structure determined by \citetalias{Baek2020}, which assumes spherical symmetry, and we ignore the conical outflow required by their SED fitting to recreate the near infrared emission. In reality, protostellar envelopes can be flattened by the effects of self-gravity and rotation, and they may contain substructures such as protostellar disks and outflow cavities, all of which influence the observed light curve. Full 3D modeling of the envelope would allow an exploration of these effects and their possible application to observations. Such ``reverberation mapping'' is often used to probe the structure of accretion disks surrounding highly variable active galactic nuclei \citep{Peterson1993}, and has also been employed to measure the radius of an inner dust hole in a protoplanetary disk by comparing the time delay of optical accretion variability with the near infrared response of the dust disk \citep{Meng2016}. When modeling the emission, we assume that the envelope is optically thin throughout, which is reasonable for submm/mm observations but breaks down for shorter wavelengths. Furthermore, the accretion disk around the embedded protostar is possibly bright and optically thick at submm/mm wavelengths \citep{Galvan-Madrid2018,Li2017}. Given the beam sizes explored in this paper (Table \ref{tab:model_telescopes}), the disk should play only a small role in the brightness variations, however, with higher angular resolution the disk would dominate, both complicating the modeling and providing a potential unique constraint on the disk physical properties.

We now consider the agreement between our EC 53 model and the observational constraints. The free parameters of our fiducial burst model are the date and value of the maximum luminosity, the floor temperature, and the outburst period, the fitted values of which are listed in Table \ref{tab:ec53_fiducial}. The minimum and maximum luminosity in our fiducial model are 3.6 $L_\odot$ and 17 $L_\odot$, which is quite similar to the same values used by \citetalias{Baek2020} of 4 $L_\odot$ and 17 $L_\odot$ to fit the quiescent and outbursting SED of EC 53 when the effects of heating by the ISRF are included. 

Our fiducial floor temperature of $T_\mathrm{floor}$ = 24 K is somewhat higher than in the radiative transfer models of \citetalias{Baek2020}, where the temperature at 10000 au is $\sim17$ K. As shown in Figure \ref{fig:envelope_comp}, the amplitude of the JCMT 850 $\mu$m light curve is particularly sensitive to the value of $T_\mathrm{floor}$ in our models. The discrepancy is likely due to the difference in the manner in which we apply $T_\mathrm{floor}$; for simplicity our calculations fix a lower temperature threshold  rather than properly calculating the local temperature due to both the incoming ISRF and the outgoing accretion luminosity. Alternatively, the outer envelope may have a steeper radial density power-law, $\alpha \sim -2$, which would also reduce the contribution of the larger scales probed by the JCMT.

Despite the simplifications, the burst model predicts amplitudes for the SMA and JCMT 850 $\mu$m light curves that are good fits to the observations (see Figure \ref{fig:ec_53_fid_model_fit}). We have also produced model light curves for the parameters of the ACA and JCMT 450 $\mu$m observations (Table \ref{tab:model_telescopes}). Comparison of the model and observed ACA light curves is significantly complicated by the sparsity of data during the outbursts of EC 53 (Figure \ref{fig:aca_lc}). Comparing the late 2019 minimum and Summer 2021 maximum flux at the ACA suggests an outburst amplitude of $\sim 1.6$, whereas our model predicts an amplitude of only 1.35. There is clear intrinsic variability in the strength of the EC 53 outbursts \citep{Lee2020yh}, however, and thus fitting to measurements spread over two burst cycles in the comparison is questionable. Interestingly, our model for the JCMT 450 $\mu$m observations predicts an amplitude of $\sim 1.4$, which is stronger than the 850 $\mu$m amplitude of $\sim 1.3$ but weaker than the estimate from the 450 $\mu$m observations of the 2021 outburst of $\sim 1.6$. Two caveats must be noted, however. First, like the ACA observations, the JCMT 450 $\mu$m observations are relatively sparse across the 2021 outburst, making it more difficult to compare directly than is the case for the SMA and JCMT 850 $\mu$m observations. Second, the beam shape of the JCMT 450 $\mu$m observations is more complex than the simple Gaussian beam in our model, and includes an additional contribution from a larger 40\arcsec\ error beam \citep{difrancesco2008,mairs2021} which adds additional modulations.

\subsection{Variable Molecular Line Emission}

In this work, we have focused entirely on the response of the thermal dust continuum to accretion bursts; however, a variety of submm/mm molecular emission lines may also be sensitive to thermal and chemical changes in the circumstellar environment. At envelope scales, \citet{Johnstone2013} showed that the equilibrium time for the molecular gas component is significantly longer than the dust response but as one probes the smallest scales, the highest densities, and the highest temperatures this imbalance diminishes.  Thus, with higher resolution observations, transitions of CO and its isotopologues may trace variations in gas temperature, while CH$_3$OH, SO, SO$_2$, and SiO transitions might trace warm gas and shocked material affected by the outburst. Furthermore, HCO+, CN, and HCN respond to increases in X-ray and UV luminosity in the accretion shock potentially triggered by the outburst. Indeed, variable H$^{13}$CO$^+$ J=3--2 emission possibly connected with an X-ray flare from a magnetic reconnection event has been detected in ALMA observations of a T Tauri star protoplanetary disk \citep{Cleeves2017}. Our SMA and ACA monitoring observations have detected a variety of these lines, which we plan to analyze for indications of variability in a forthcoming paper.

\section{Conclusions}
\label{sec:conc}

In this paper, we have interpreted contemporaneous sub-mm/mm light curves of variable protostars as observed by the ALMA ACA, SMA, and JCMT using a toy model of the envelope dust response to accretion luminosity variations. Our major results are as follows:

\begin{itemize}
    \item Relative flux calibration is vital to this work, and accurate to about 3\% for ACA/SMA (Section \ref{sec:dr}). This is a significant improvement over the typical flux calibration accuracy of sub-mm/mm interferometers of 10-15\%.
    \item We have robustly detected variability in our ACA observations of EC 53 (V371 Ser) and SMM 10, two known submm variables based on JCMT Transient Survey monitoring (Figure \ref{fig:aca_lc}). 
    \item A delay between the peak amplitude of the EC 53 (V371 Ser) outburst at the SMA and JCMT is seen, and the amplitude of the SMA outburst is somewhat stronger (Figure \ref{fig:sma_lc}). 
    \item We have developed a toy model of the envelope response in EC 53 to show that the delay and difference in amplitude between the SMA and JCMT (850 \micron) are plausibly explained by 1: the light travel time delay through the envelope; and 2: the dilution of the envelope response at the JCMT by the incorporation of more cold envelope material in the beam (Section \ref{sec:toy_model}). The JCMT amplitude is particularly sensitive to the heating by the ISRF, described by an envelope floor temperature in our model.
    \item We have further explored the effects a variety of bursts and observational properties of our toy model (Section \ref{sec:further_modeling}), and shown that high frequency bursts can be washed out by the lag in the response with a larger beam, which may be ocurring for SMM 10. A stronger envelope response tied more closely to luminosity variations is expected at shorter far infrared wavelengths. 
\end{itemize}


The authors are grateful for the excellent support provided by the ALMA and SMA staff for our observing programs. We would particularly like to thank David Wilner, Mark Gurwell, and Charlie Qi for useful discussions on the planning and execution of our SMA observations.

The authors wish to recognize and acknowledge the very significant cultural role and reverence that the summit of Maunakea has always had within the indigenous Hawaiian community. We are most fortunate to have the opportunity to conduct observations from this mountain. 

The James Clerk Maxwell Telescope is operated by the East Asian Observatory on behalf of The National Astronomical Observatory of Japan; Academia Sinica Institute of Astronomy and Astrophysics; the Korea Astronomy and Space Science Institute; the Operation, Maintenance and Upgrading Fund for Astronomical Telescopes and Facility Instruments, budgeted from the Ministry of Finance (MOF) of China and administrated by the Chinese Academy of Sciences (CAS), as well as the National Key R\&D Program of China (No. 2017YFA0402700). Additional funding support is provided by the Science and Technology Facilities Council of the United Kingdom and participating universities in the United Kingdom and Canada. Additional funds for the construction of SCUBA-2 were provided by the Canada Foundation for Innovation. The James Clerk Maxwell Telescope has historically been operated by the Joint Astronomy Centre on behalf of the Science and Technology Facilities Council of the United Kingdom, the National Research Council (NRC) of Canada and the Netherlands Organisation for Scientific Research. The JCMT Transient Survey project codes are M16AL001 and M20AL007.

This paper makes use of the following ALMA data: ADS/JAO.ALMA\#2018.1.00917.S, ADS/JAO.ALMA\#2019.1.00475.S
ALMA is a partnership of ESO (representing its member states), NSF (USA) and NINS (Japan), together with NRC (Canada), NSC and ASIAA (Taiwan), and KASI (Republic of Korea), in cooperation with the Republic of Chile.  The Joint ALMA Observatory is operated by ESO, AUI/NRAO and NAOJ. The National Radio Astronomy Observatory is a facility of the National Science Foundation  operated under cooperative agreement by Associated Universities, Inc.

The Submillimeter Array is a joint project between the Smithsonian Astrophysical Observatory and the Academia Sinica Institute of Astronomy and Astrophysics and is funded by the Smithsonian Institution and the Academia Sinica.

This research used the facilities of the Canadian Astronomy Data Centre operated by NRC Canada with the support of the Canadian Space Agency. D.J.\ is supported by NRC Canada and by an NSERC Discovery Grant. G.J.H.\ is supported by general grant 12173003 awarded by the National Science Foundation of China.


\bibliography{paper}{}

\begin{thebibliography}{}
\expandafter\ifx\csname natexlab\endcsname\relax\def\natexlab#1{#1}\fi
\providecommand{\url}[1]{\href{#1}{#1}}
\providecommand{\dodoi}[1]{doi:~\href{http://doi.org/#1}{\nolinkurl{#1}}}
\providecommand{\doeprint}[1]{\href{http://ascl.net/#1}{\nolinkurl{http://ascl.net/#1}}}
\providecommand{\doarXiv}[1]{\href{https://arxiv.org/abs/#1}{\nolinkurl{https://arxiv.org/abs/#1}}}

\bibitem[{{Akiyama} {et~al.}(2017){Akiyama}, {Kuramochi}, {Ikeda}, {Fish},
  {Tazaki}, {Honma}, {Doeleman}, {Broderick}, {Dexter}, {Mo{\'s}cibrodzka},
  {Bouman}, {Chael}, \& {Zaizen}}]{Akiyama2017}
{Akiyama}, K., {Kuramochi}, K., {Ikeda}, S., {et~al.} 2017, \apj, 838, 1,
  \dodoi{10.3847/1538-4357/aa6305}

\bibitem[{{Andr{\'e}} {et~al.}(2019){Andr{\'e}}, {Hughes}, {Guillet},
  {Boulanger}, {Bracco}, {Ntormousi}, {Arzoumanian}, {Maury}, {Bernard},
  {Bontemps}, {Ristorcelli}, {Girart}, {Motte}, {Tassis}, {Pantin},
  {Montmerle}, {Johnstone}, {Gabici}, {Efstathiou}, {Basu}, {B{\'e}thermin},
  {Beuther}, {Braine}, {Francesco}, {Falgarone}, {Ferri{\`e}re}, {Fletcher},
  {Galametz}, {Giard}, {Hennebelle}, {Jones}, {Kepley}, {Kwon}, {Lagache},
  {Lesaffre}, {Levrier}, {Li}, {Li}, {Mao}, {Nakagawa}, {Onaka}, {Paladino},
  {Peretto}, {Poglitsch}, {Rev{\'e}ret}, {Rodriguez}, {Sauvage}, {Soler},
  {Spinoglio}, {Tabatabaei}, {Tritsis}, {van der Tak}, {Ward-Thompson},
  {Wiesemeyer}, {Ysard}, \& {Zhang}}]{andre2019}
{Andr{\'e}}, P., {Hughes}, A., {Guillet}, V., {et~al.} 2019, \pasa, 36, e029,
  \dodoi{10.1017/pasa.2019.20}

\bibitem[{{Armitage}(2016)}]{armitage2016}
{Armitage}, P.~J. 2016, \apjl, 833, L15, \dodoi{10.3847/2041-8213/833/2/L15}

\bibitem[{{Baek} {et~al.}(2020){Baek}, {MacFarlane}, {Lee}, {Stamatellos},
  {Herczeg}, {Johnstone}, {Pe{\~n}a}, {Varricatt}, {Hodapp}, {Chen}, \&
  {Kang}}]{Baek2020}
{Baek}, G., {MacFarlane}, B.~A., {Lee}, J.-E., {et~al.} 2020, \apj, 895, 27,
  \dodoi{10.3847/1538-4357/ab8ad4}

\bibitem[{{CCAT-Prime collaboration} {et~al.}(2021){CCAT-Prime collaboration},
  {Aravena}, {Austermann}, {Basu}, {Battaglia}, {Beringue}, {Bertoldi},
  {Bigiel}, {Bond}, {Breysse}, {Broughton}, {Bustos}, {Chapman}, {Charmetant},
  {Choi}, {Chung}, {Clark}, {Cothard}, {Crites}, {Dev}, {Douglas}, {Duell},
  {Ebina}, {Erler}, {Fich}, {Fissel}, {Foreman}, {Gao}, {Garc{\'\i}a},
  {Giovanelli}, {Haynes}, {Hensley}, {Herter}, {Higgins}, {Huber}, {Hubmayr},
  {Johnstone}, {Karoumpis}, {Keating}, {Komatsu}, {Li}, {Magnelli}, {Matthews},
  {Meerburg}, {Meyers}, {Muralidhara}, {Murray}, {Niemack}, {Nikola}, {Okada},
  {Riechers}, {Rosolowsky}, {Roy}, {Sadavoy}, {Schaaf}, {Schilke}, {Scott},
  {Simon}, {Sinclair}, {Sivakoff}, {Stacey}, {Stutz}, {Stutzki}, {Tahani},
  {Thanjavur}, {Timmermann}, {Ullom}, {van Engelen}, {Vavagiakis}, {Vissers},
  {Wheeler}, {White}, {Zhu}, \& {Zou}}]{ccat2021}
{CCAT-Prime collaboration}, {Aravena}, M., {Austermann}, J.~E., {et~al.} 2021,
  arXiv e-prints, arXiv:2107.10364.
\newblock \doarXiv{2107.10364}

\bibitem[{{Chael} {et~al.}(2018){Chael}, {Johnson}, {Bouman}, {Blackburn},
  {Akiyama}, \& {Narayan}}]{Chael2018}
{Chael}, A.~A., {Johnson}, M.~D., {Bouman}, K.~L., {et~al.} 2018, \apj, 857,
  23, \dodoi{10.3847/1538-4357/aab6a8}

\bibitem[{{Chen} {et~al.}(2021){Chen}, {Sun}, {Chini}, {Haas}, {Jiang}, \&
  {Chen}}]{Chen2021}
{Chen}, Z., {Sun}, W., {Chini}, R., {et~al.} 2021, \apj, 922, 90,
  \dodoi{10.3847/1538-4357/ac2151}

\bibitem[{{Cleeves} {et~al.}(2017){Cleeves}, {Bergin}, {{\"O}berg}, {Andrews},
  {Wilner}, \& {Loomis}}]{Cleeves2017}
{Cleeves}, L.~I., {Bergin}, E.~A., {{\"O}berg}, K.~I., {et~al.} 2017, \apjl,
  843, L3, \dodoi{10.3847/2041-8213/aa76e2}

\bibitem[{{Contreras Pe{\~n}a} {et~al.}(2020){Contreras Pe{\~n}a}, {Johnstone},
  {Baek}, {Herczeg}, {Mairs}, {Scholz}, {Lee}, \& {JCMT Transient
  Team}}]{ContrerasPena2020}
{Contreras Pe{\~n}a}, C., {Johnstone}, D., {Baek}, G., {et~al.} 2020, \mnras,
  495, 3614, \dodoi{10.1093/mnras/staa1254}

\bibitem[{{Cook} {et~al.}(2019){Cook}, {Tobin}, {Skrutskie}, \&
  {Nelson}}]{Cook2019}
{Cook}, B.~T., {Tobin}, J.~J., {Skrutskie}, M.~F., \& {Nelson}, M.~J. 2019,
  \aap, 626, A51, \dodoi{10.1051/0004-6361/201935419}

\bibitem[{{Dahm} \& {Hillenbrand}(2020)}]{Dahm2020}
{Dahm}, S.~E., \& {Hillenbrand}, L.~A. 2020, \aj, 160, 278,
  \dodoi{10.3847/1538-3881/abbfa2}

\bibitem[{{D'Angelo} \& {Spruit}(2012)}]{dangelo2012}
{D'Angelo}, C.~R., \& {Spruit}, H.~C. 2012, \mnras, 420, 416,
  \dodoi{10.1111/j.1365-2966.2011.20046.x}

\bibitem[{{Dempsey} {et~al.}(2013){Dempsey}, {Friberg}, {Jenness}, {Tilanus},
  {Thomas}, {Holland}, {Bintley}, {Berry}, {Chapin}, {Chrysostomou}, {Davis},
  {Gibb}, {Parsons}, \& {Robson}}]{Dempsey2013}
{Dempsey}, J.~T., {Friberg}, P., {Jenness}, T., {et~al.} 2013, \mnras, 430,
  2534, \dodoi{10.1093/mnras/stt090}

\bibitem[{{Di Francesco} {et~al.}(2008){Di Francesco}, {Johnstone}, {Kirk},
  {MacKenzie}, \& {Ledwosinska}}]{difrancesco2008}
{Di Francesco}, J., {Johnstone}, D., {Kirk}, H., {MacKenzie}, T., \&
  {Ledwosinska}, E. 2008, \apjs, 175, 277, \dodoi{10.1086/523645}

\bibitem[{{Fischer} {et~al.}(2019){Fischer}, {Dunham}, {Green}, {Hatchell},
  {Johnstone}, {Battersby}, {Klaassen}, {Li}, {Offner}, {Pontoppidan},
  {Sewilo}, {Stephens}, {Tobin}, {Brogan}, {Gutermuth}, {Looney}, {Megeath},
  {Padgett}, \& {Roellig}}]{fischer2019}
{Fischer}, W., {Dunham}, M., {Green}, J., {et~al.} 2019, \baas, 51, 495.
\newblock \doarXiv{1903.07628}

\bibitem[{{Fischer} {et~al.}(2022){Fischer}, {Hillenbrand}, {Herczeg},
  {Johnstone}, {K{\'o}sp{\'a}l}, \& {Dunham}}]{Fischer2022}
{Fischer}, W.~J., {Hillenbrand}, L.~A., {Herczeg}, G.~J., {et~al.} 2022, arXiv
  e-prints, arXiv:2203.11257.
\newblock \doarXiv{2203.11257}

\bibitem[{{Francis} {et~al.}(2020){Francis}, {Johnstone}, {Herczeg}, {Hunter},
  \& {Harsono}}]{Francis2020}
{Francis}, L., {Johnstone}, D., {Herczeg}, G., {Hunter}, T.~R., \& {Harsono},
  D. 2020, \aj, 160, 270, \dodoi{10.3847/1538-3881/abbe1a}

\bibitem[{{Galv{\'a}n-Madrid} {et~al.}(2018){Galv{\'a}n-Madrid}, {Liu},
  {Izquierdo}, {Miotello}, {Zhao}, {Carrasco-Gonz{\'a}lez}, {Lizano}, \&
  {Rodr{\'\i}guez}}]{Galvan-Madrid2018}
{Galv{\'a}n-Madrid}, R., {Liu}, H.~B., {Izquierdo}, A.~F., {et~al.} 2018, \apj,
  868, 39, \dodoi{10.3847/1538-4357/aae779}

\bibitem[{{Guo} {et~al.}(2022){Guo}, {Lucas}, {Smith}, {Clarke}, {Contreras
  Pe{\~n}a}, {Bayo}, {Brice{\~n}o}, {Elias}, {Kurtev}, {Borissova},
  {Alonso-Garc{\'\i}a}, {Minniti}, {Catelan}, {Nikzat}, {Morris}, \&
  {Miller}}]{Guo2022}
{Guo}, Z., {Lucas}, P.~W., {Smith}, L.~C., {et~al.} 2022, \mnras, 513, 1015,
  \dodoi{10.1093/mnras/stac768}

\bibitem[{{Hartmann} {et~al.}(2016){Hartmann}, {Herczeg}, \&
  {Calvet}}]{Hartmann2016}
{Hartmann}, L., {Herczeg}, G., \& {Calvet}, N. 2016, \araa, 54, 135,
  \dodoi{10.1146/annurev-astro-081915-023347}

\bibitem[{{Herczeg} {et~al.}(2017){Herczeg}, {Johnstone}, {Mairs}, {Hatchell},
  {Lee}, {Bower}, {Chen}, {Aikawa}, {Yoo}, {Kang}, {Kang}, {Chen}, {Williams},
  {Bae}, {Dunham}, {Vorobyov}, {Zhu}, {Rao}, {Kirk}, {Takahashi}, {Morata},
  {Lacaille}, {Lane}, {Pon}, {Scholz}, {Samal}, {Bell}, {Graves}, {Lee},
  {Parsons}, {He}, {Zhou}, {Kim}, {Chapman}, {Drabek-Maunder}, {Chung},
  {Eyres}, {Forbrich}, {Hillenbrand}, {Inutsuka}, {Kim}, {Kim}, {Kuan}, {Kwon},
  {Lai}, {Lalchand}, {Lee}, {Lee}, {Long}, {Lyo}, {Qian}, {Scicluna}, {Soam},
  {Stamatellos}, {Takakuwa}, {Tang}, {Wang}, \& {Wang}}]{Herczeg2017}
{Herczeg}, G.~J., {Johnstone}, D., {Mairs}, S., {et~al.} 2017, \apj, 849, 43,
  \dodoi{10.3847/1538-4357/aa8b62}

\bibitem[{{Herczeg} {et~al.}(2019){Herczeg}, {Kuhn}, {Zhou}, {Hatchell},
  {Manara}, {Johnstone}, {Dunham}, {Bhardwaj}, {Jose}, \& {Yuan}}]{herczeg2019}
{Herczeg}, G.~J., {Kuhn}, M.~A., {Zhou}, X., {et~al.} 2019, \apj, 878, 111,
  \dodoi{10.3847/1538-4357/ab1d67}

\bibitem[{{Hillenbrand} {et~al.}(2019){Hillenbrand}, {Reipurth}, {Connelley},
  {Cutri}, \& {Isaacson}}]{hillenbrand2019}
{Hillenbrand}, L.~A., {Reipurth}, B., {Connelley}, M., {Cutri}, R.~M., \&
  {Isaacson}, H. 2019, \aj, 158, 240, \dodoi{10.3847/1538-3881/ab4e16}

\bibitem[{{Hillenbrand} {et~al.}(2018){Hillenbrand}, {Contreras Pe{\~n}a},
  {Morrell}, {Naylor}, {Kuhn}, {Cutri}, {Rebull}, {Hodgkin}, {Froebrich}, \&
  {Mainzer}}]{hillenbrand2018}
{Hillenbrand}, L.~A., {Contreras Pe{\~n}a}, C., {Morrell}, S., {et~al.} 2018,
  \apj, 869, 146, \dodoi{10.3847/1538-4357/aaf414}

\bibitem[{{Hodapp} \& {Chini}(2015)}]{Hodapp2015}
{Hodapp}, K.~W., \& {Chini}, R. 2015, \apj, 813, 107,
  \dodoi{10.1088/0004-637X/813/2/107}

\bibitem[{{Holland} {et~al.}(2013){Holland}, {Bintley}, {Chapin},
  {Chrysostomou}, {Davis}, {Dempsey}, {Duncan}, {Fich}, {Friberg}, {Halpern},
  {Irwin}, {Jenness}, {Kelly}, {MacIntosh}, {Robson}, {Scott}, {Ade},
  {Atad-Ettedgui}, {Berry}, {Craig}, {Gao}, {Gibb}, {Hilton}, {Hollister},
  {Kycia}, {Lunney}, {McGregor}, {Montgomery}, {Parkes}, {Tilanus}, {Ullom},
  {Walther}, {Walton}, {Woodcraft}, {Amiri}, {Atkinson}, {Burger}, {Chuter},
  {Coulson}, {Doriese}, {Dunare}, {Economou}, {Niemack}, {Parsons},
  {Reintsema}, {Sibthorpe}, {Smail}, {Sudiwala}, \& {Thomas}}]{Holland2013}
{Holland}, W.~S., {Bintley}, D., {Chapin}, E.~L., {et~al.} 2013, \mnras, 430,
  2513, \dodoi{10.1093/mnras/sts612}

\bibitem[{{Honma} {et~al.}(2014){Honma}, {Akiyama}, {Uemura}, \&
  {Ikeda}}]{Honma2014}
{Honma}, M., {Akiyama}, K., {Uemura}, M., \& {Ikeda}, S. 2014, \pasj, 66, 95,
  \dodoi{10.1093/pasj/psu070}

\bibitem[{{Hunter} {et~al.}(2006){Hunter}, {Brogan}, {Megeath}, {Menten},
  {Beuther}, \& {Thorwirth}}]{Hunter2006}
{Hunter}, T.~R., {Brogan}, C.~L., {Megeath}, S.~T., {et~al.} 2006, \apj, 649,
  888, \dodoi{10.1086/505965}

\bibitem[{{Hunter} {et~al.}(2017){Hunter}, {Brogan}, {MacLeod}, {Cyganowski},
  {Chandler}, {Chibueze}, {Friesen}, {Indebetouw}, {Thesner}, \&
  {Young}}]{Hunter2017}
{Hunter}, T.~R., {Brogan}, C.~L., {MacLeod}, G., {et~al.} 2017, \apjl, 837,
  L29, \dodoi{10.3847/2041-8213/aa5d0e}

\bibitem[{{Jhan} {et~al.}(2022){Jhan}, {Lee}, {Johnstone}, {Liu}, {Liu},
  {Hirano}, {Tatematsu}, {Dutta}, {Moraghan}, {Shang}, {Lee}, {Li}, {Liu},
  {Hsu}, {Kwon}, {Sahu}, {Liu}, {Kim}, {Luo}, {Qin}, {Sanhueza}, {Bronfman},
  {Qizhou}, {Eden}, {Traficante}, {Lee}, \& {Almasop Team}}]{Jhan2022}
{Jhan}, K.-S., {Lee}, C.-F., {Johnstone}, D., {et~al.} 2022, \apjl, 931, L5,
  \dodoi{10.3847/2041-8213/ac6a53}

\bibitem[{{Johnstone} {et~al.}(2013){Johnstone}, {Hendricks}, {Herczeg}, \&
  {Bruderer}}]{Johnstone2013}
{Johnstone}, D., {Hendricks}, B., {Herczeg}, G.~J., \& {Bruderer}, S. 2013,
  \apj, 765, 133, \dodoi{10.1088/0004-637X/765/2/133}

\bibitem[{{Johnstone} {et~al.}(2018){Johnstone}, {Herczeg}, {Mairs},
  {Hatchell}, {Bower}, {Kirk}, {Lane}, {Bell}, {Graves}, {Aikawa}, {Chen},
  {Chen}, {Kang}, {Kang}, {Lee}, {Morata}, {Pon}, {Scicluna}, {Scholz},
  {Takahashi}, {Yoo}, \& {JCMT Transient Team}}]{Johnstone2018}
{Johnstone}, D., {Herczeg}, G.~J., {Mairs}, S., {et~al.} 2018, \apj, 854, 31,
  \dodoi{10.3847/1538-4357/aaa764}

\bibitem[{{J{\o}rgensen} {et~al.}(2013){J{\o}rgensen}, {Visser}, {Sakai},
  {Bergin}, {Brinch}, {Harsono}, {Lindberg}, {van Dishoeck}, {Yamamoto},
  {Bisschop}, \& {Persson}}]{Jorgensen2013}
{J{\o}rgensen}, J.~K., {Visser}, R., {Sakai}, N., {et~al.} 2013, \apjl, 779,
  L22, \dodoi{10.1088/2041-8205/779/2/L22}

\bibitem[{{Kackley} {et~al.}(2010){Kackley}, {Scott}, {Chapin}, \&
  {Friberg}}]{kackley2010}
{Kackley}, R., {Scott}, D., {Chapin}, E., \& {Friberg}, P. 2010, in Proc. SPIE,
  ed. T.~G. {Phillips} \& J.~{Zmuidzinas}, Vol. 7740, 1. SPIE, Bellingham, WA,
  \dodoi{10.1117/12.857397}

\bibitem[{{Lee} {et~al.}(2020){Lee}, {Johnstone}, {Lee}, {Herczeg}, {Mairs},
  {Varricatt}, {Hodapp}, {Naylor}, {Pe{\~n}a}, {Baek}, {Haas}, {Chini}, \&
  {JCMT Transient Team}}]{Lee2020yh}
{Lee}, Y.-H., {Johnstone}, D., {Lee}, J.-E., {et~al.} 2020, \apj, 903, 5,
  \dodoi{10.3847/1538-4357/abb6fe}

\bibitem[{{Lee} {et~al.}(2021){Lee}, {Johnstone}, {Lee}, {Herczeg}, {Mairs},
  {Contreras-Pe{\~n}a}, {Hatchell}, {Naylor}, {Bell}, {Bourke}, {Broughton},
  {Francis}, {Gupta}, {Harsono}, {Liu}, {Park}, {Plovie}, {Moriarty-Schieven},
  {Scholz}, {Sharma}, {Stella Teixeira}, {Wang}, {Aikawa}, {Bower}, {Vivien
  Chen}, {Bae}, {Baek}, {Chapman}, {Ping Chen}, {Du}, {Dutta}, {Forbrich},
  {Guo}, {Inutsuka}, {Kang}, {Kirk}, {Kuan}, {Kwon}, {Lai}, {Lalchand}, {Lane},
  {Lee}, {Liu}, {Morata}, {Pearson}, {Pon}, {Sahu}, {Shang}, {Stamatellos},
  {Tang}, {Xu}, {Yoo}, \& {Rawlings}}]{Lee2021}
---. 2021, \apj, 920, 119, \dodoi{10.3847/1538-4357/ac1679}

\bibitem[{{Li} {et~al.}(2017){Li}, {Liu}, {Hasegawa}, \& {Hirano}}]{Li2017}
{Li}, J. I.-H., {Liu}, H.~B., {Hasegawa}, Y., \& {Hirano}, N. 2017, \apj, 840,
  72, \dodoi{10.3847/1538-4357/aa6f04}

\bibitem[{{MacFarlane} {et~al.}(2019{\natexlab{a}}){MacFarlane}, {Stamatellos},
  {Johnstone}, {Herczeg}, {Baek}, {Chen}, {Kang}, \& {Lee}}]{MacFarlane2019a}
{MacFarlane}, B., {Stamatellos}, D., {Johnstone}, D., {et~al.}
  2019{\natexlab{a}}, \mnras, 487, 4465, \dodoi{10.1093/mnras/stz1570}

\bibitem[{{MacFarlane} {et~al.}(2019{\natexlab{b}}){MacFarlane}, {Stamatellos},
  {Johnstone}, {Herczeg}, {Baek}, {Chen}, {Kang}, \& {Lee}}]{MacFarlane2019b}
---. 2019{\natexlab{b}}, \mnras, 487, 5106, \dodoi{10.1093/mnras/stz1512}

\bibitem[{{Mainzer} {et~al.}(2011){Mainzer}, {Bauer}, {Grav}, {Masiero},
  {Cutri}, {Dailey}, {Eisenhardt}, {McMillan}, {Wright}, {Walker}, {Jedicke},
  {Spahr}, {Tholen}, {Alles}, {Beck}, {Brandenburg}, {Conrow}, {Evans},
  {Fowler}, {Jarrett}, {Marsh}, {Masci}, {McCallon}, {Wheelock}, {Wittman},
  {Wyatt}, {DeBaun}, {Elliott}, {Elsbury}, {Gautier}, {Gomillion}, {Leisawitz},
  {Maleszewski}, {Micheli}, \& {Wilkins}}]{Mainzer2011}
{Mainzer}, A., {Bauer}, J., {Grav}, T., {et~al.} 2011, \apj, 731, 53,
  \dodoi{10.1088/0004-637X/731/1/53}

\bibitem[{{Mairs} {et~al.}(2017){Mairs}, {Lane}, {Johnstone}, {Kirk},
  {Lacaille}, {Bower}, {Bell}, {Graves}, {Chapman}, \& {JCMT Transient
  Team}}]{Mairs2017Cal}
{Mairs}, S., {Lane}, J., {Johnstone}, D., {et~al.} 2017, \apj, 843, 55,
  \dodoi{10.3847/1538-4357/aa7844}

\bibitem[{{Mairs} {et~al.}(2021){Mairs}, {Dempsey}, {Bell}, {Parsons},
  {Currie}, {Friberg}, {Jiang}, {Tetarenko}, {Bintley}, {Cookson}, {Li},
  {Rawlings}, {Wouterloot}, {Berry}, {Graves}, {Mizuno}, {Acohido}, {Clark},
  {Cox}, {Fuchs}, {Hoge}, {Kemp}, {Lee}, {Matulonis}, {Montgomerie}, {Silva},
  \& {Smith}}]{mairs2021}
{Mairs}, S., {Dempsey}, J.~T., {Bell}, G.~S., {et~al.} 2021, \aj, 162, 191,
  \dodoi{10.3847/1538-3881/ac18bf}

\bibitem[{{McKee} \& {Offner}(2011)}]{McKee2011}
{McKee}, C.~F., \& {Offner}, S. R.~R. 2011, in Computational Star Formation,
  ed. J.~{Alves}, B.~G. {Elmegreen}, J.~M. {Girart}, \& V.~{Trimble}, Vol. 270,
  73--80, \dodoi{10.1017/S1743921311000202}

\bibitem[{{McMullin} {et~al.}(2007){McMullin}, {Waters}, {Schiebel}, {Young},
  \& {Golap}}]{McMullin2007}
{McMullin}, J.~P., {Waters}, B., {Schiebel}, D., {Young}, W., \& {Golap}, K.
  2007, in Astronomical Society of the Pacific Conference Series, Vol. 376,
  Astronomical Data Analysis Software and Systems XVI, ed. R.~A. {Shaw},
  F.~{Hill}, \& D.~J. {Bell}, 127

\bibitem[{{Meng} {et~al.}(2016){Meng}, {Plavchan}, {Rieke}, {Cody}, {G{\"u}th},
  {Stauffer}, {Covey}, {Carey}, {Ciardi}, {Duran-Rojas}, {Gutermuth},
  {Morales-Calder{\'o}n}, {Rebull}, \& {Watson}}]{Meng2016}
{Meng}, H. Y.~A., {Plavchan}, P., {Rieke}, G.~H., {et~al.} 2016, \apj, 823, 58,
  \dodoi{10.3847/0004-637X/823/1/58}

\bibitem[{{Muzerolle} {et~al.}(2013){Muzerolle}, {Furlan}, {Flaherty}, {Balog},
  \& {Gutermuth}}]{Muzerolle2013}
{Muzerolle}, J., {Furlan}, E., {Flaherty}, K., {Balog}, Z., \& {Gutermuth}, R.
  2013, \nat, 493, 378, \dodoi{10.1038/nature11746}

\bibitem[{{Ortiz-Le{\'o}n} {et~al.}(2017){Ortiz-Le{\'o}n}, {Dzib}, {Kounkel},
  {Loinard}, {Mioduszewski}, {Rodr{\'\i}guez}, {Torres}, {Pech}, {Rivera},
  {Hartmann}, {Boden}, {Evans}, {Brice{\~n}o}, {Tobin}, \&
  {Galli}}]{Ortiz-Leon2017}
{Ortiz-Le{\'o}n}, G.~N., {Dzib}, S.~A., {Kounkel}, M.~A., {et~al.} 2017, \apj,
  834, 143, \dodoi{10.3847/1538-4357/834/2/143}

\bibitem[{{Park} {et~al.}(2021){Park}, {Lee}, {Contreras Pe{\~n}a},
  {Johnstone}, {Herczeg}, {Lee}, {Lee}, {Bhardwaj}, \&
  {Moriarty-Schieven}}]{Park2021}
{Park}, W., {Lee}, J.-E., {Contreras Pe{\~n}a}, C., {et~al.} 2021, \apj, 920,
  132, \dodoi{10.3847/1538-4357/ac1745}

\bibitem[{{Peterson}(1993)}]{Peterson1993}
{Peterson}, B.~M. 1993, \pasp, 105, 247, \dodoi{10.1086/133140}

\bibitem[{{Plunkett} {et~al.}(2015){Plunkett}, {Arce}, {Mardones}, {van
  Dokkum}, {Dunham}, {Fern{\'a}ndez-L{\'o}pez}, {Gallardo}, \&
  {Corder}}]{Plunkett2015}
{Plunkett}, A.~L., {Arce}, H.~G., {Mardones}, D., {et~al.} 2015, \nat, 527, 70,
  \dodoi{10.1038/nature15702}

\bibitem[{{Primiani} {et~al.}(2016){Primiani}, {Young}, {Young}, {Patel},
  {Wilson}, {Vertatschitsch}, {Chitwood}, {Srinivasan}, {MacMahon}, \&
  {Weintroub}}]{Primiani2016}
{Primiani}, R.~A., {Young}, K.~H., {Young}, A., {et~al.} 2016, Journal of
  Astronomical Instrumentation, 5, 1641006, \dodoi{10.1142/S2251171716410063}

\bibitem[{{Rebull} {et~al.}(2014){Rebull}, {Cody}, {Covey}, {G{\"u}nther},
  {Hillenbrand}, {Plavchan}, {Poppenhaeger}, {Stauffer}, {Wolk}, {Gutermuth},
  {Morales-Calder{\'o}n}, {Song}, {Barrado}, {Bayo}, {James}, {Hora}, {Vrba},
  {Alves de Oliveira}, {Bouvier}, {Carey}, {Carpenter}, {Favata}, {Flaherty},
  {Forbrich}, {Hernandez}, {McCaughrean}, {Megeath}, {Micela}, {Smith},
  {Terebey}, {Turner}, {Allen}, {Ardila}, {Bouy}, \& {Guieu}}]{Rebull2014}
{Rebull}, L.~M., {Cody}, A.~M., {Covey}, K.~R., {et~al.} 2014, \aj, 148, 92,
  \dodoi{10.1088/0004-6256/148/5/92}

\bibitem[{{Safron} {et~al.}(2015){Safron}, {Fischer}, {Megeath}, {Furlan},
  {Stutz}, {Stanke}, {Billot}, {Rebull}, {Tobin}, {Ali}, {Allen}, {Booker},
  {Watson}, \& {Wilson}}]{Safron2015}
{Safron}, E.~J., {Fischer}, W.~J., {Megeath}, S.~T., {et~al.} 2015, \apjl, 800,
  L5, \dodoi{10.1088/2041-8205/800/1/L5}

\bibitem[{{Yoo} {et~al.}(2017){Yoo}, {Lee}, {Mairs}, {Johnstone}, {Herczeg},
  {Kang}, {Kang}, {Cho}, \& {JCMT Transient Team}}]{Yoo2017}
{Yoo}, H., {Lee}, J.-E., {Mairs}, S., {et~al.} 2017, \apj, 849, 69,
  \dodoi{10.3847/1538-4357/aa8c0a}

\bibitem[{{Zakri} {et~al.}(2022){Zakri}, {Megeath}, {Fischer}, {Gutermuth},
  {Furlan}, {Hartmann}, {Karnath}, {Osorio}, {Safron}, {Stanke}, {Stutz},
  {Tobin}, {Allen}, {Federman}, {Habel}, {Manoj}, {Narang}, {Pokhrel},
  {Rebull}, {Sheehan}, \& {Watson}}]{Zakri2022}
{Zakri}, W., {Megeath}, S.~T., {Fischer}, W.~J., {et~al.} 2022, \apjl, 924,
  L23, \dodoi{10.3847/2041-8213/ac46ae}

\end{thebibliography}
\bibliographystyle{aasjournal}



\appendix
\section{Deconvolved Deep SMA and ACA images of Targets}

\begin{figure}[htb]
    \centering
    \includegraphics[scale=0.9]{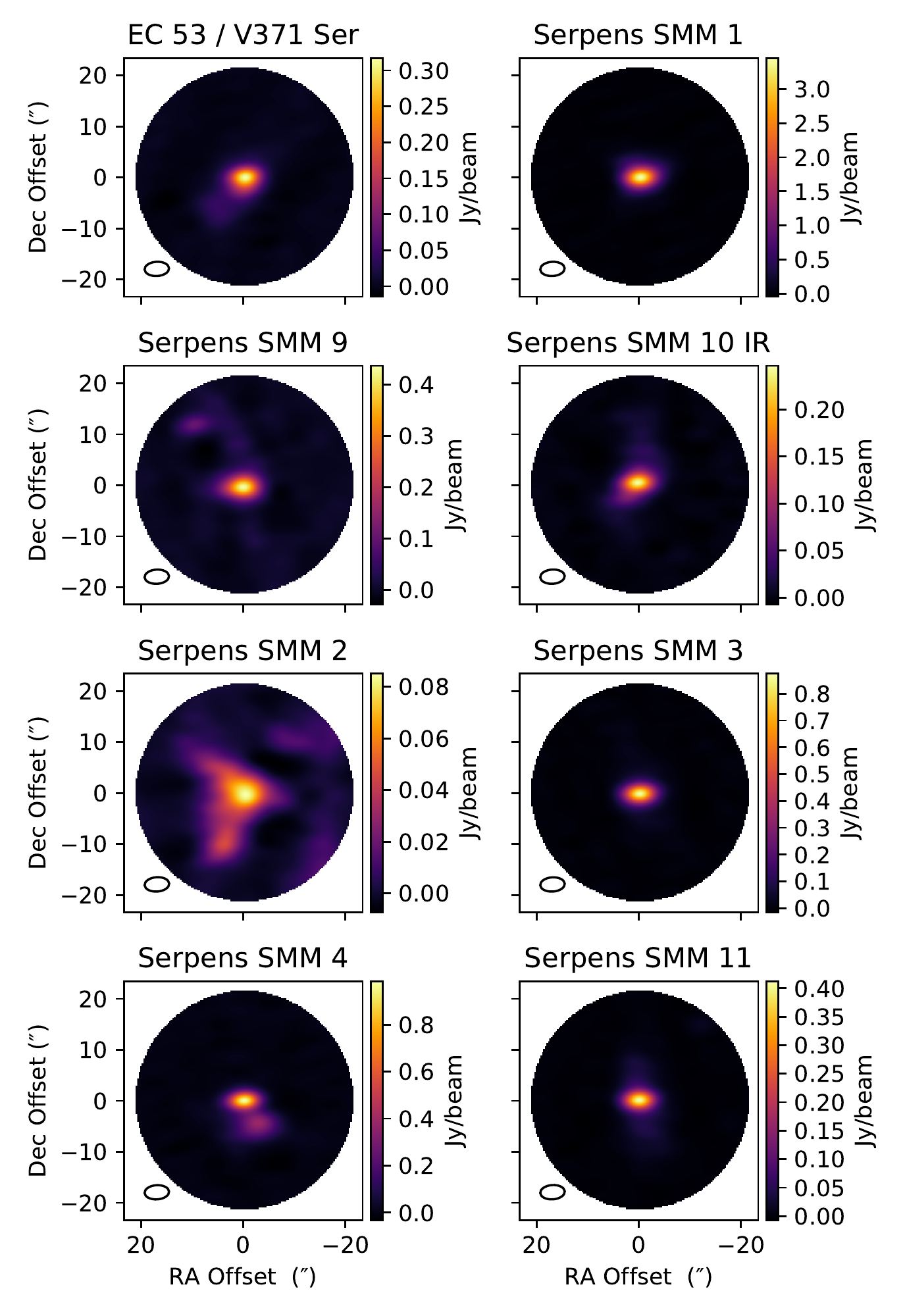}
    \caption{Gallery of sub-mm continuum images for the ALMA ACA observations, produced by concatenating data from all epochs with the relative calibration in Section \ref{sec:rel_calibration} applied. The beam size is denoted by the black ellipse in the lower left corner.}
    \label{fig:aca_deep_gallery}
\end{figure}

\begin{figure}[htb]
    \centering
    \includegraphics[width=1.0
    \textwidth]{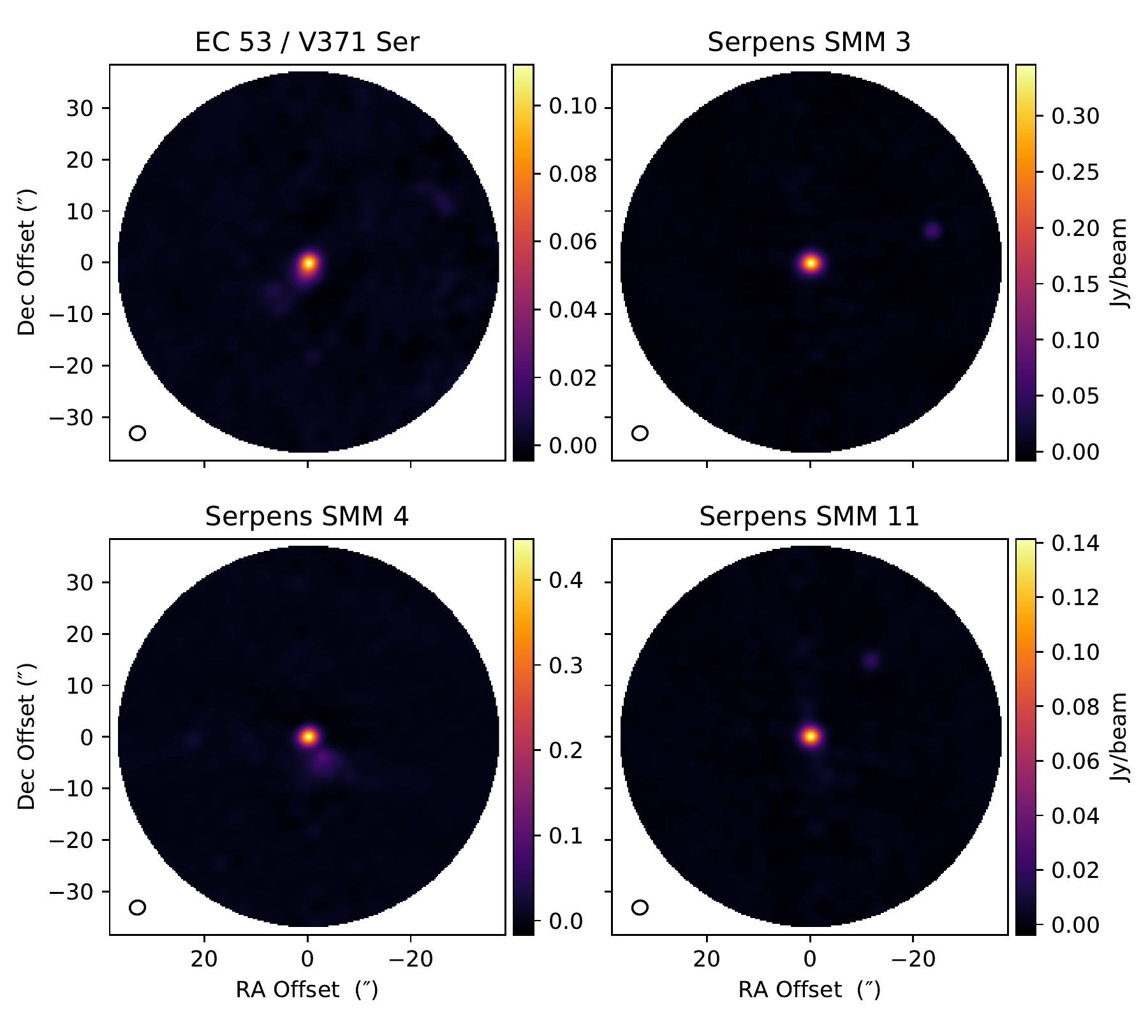}
    \caption{As Figure \ref{fig:aca_deep_gallery}, but for the SMA observed targets.}
    \label{fig:sma_deep_gallery}
\end{figure}

In Figures \ref{fig:aca_deep_gallery} and \ref{fig:sma_deep_gallery} we present the deep, ACA and SMA images of each Serpens Main source, combining all epochs and with the relative calibration in Section \ref{sec:rel_calibration} applied.

\end{document}